\documentclass[aps,prm,preprint,longbibliography]{revtex4-1}
\usepackage[utf8]{inputenc}
\usepackage{amssymb}
\usepackage{amsthm}
\usepackage{amsmath}
\usepackage{graphicx}
\usepackage{multirow}
\graphicspath{ {./} {./images/} }
\usepackage[skip=0.5\baselineskip]{caption}
\usepackage{float}
\usepackage{times}
\usepackage{xcolor}
\usepackage[sort&compress]{natbib}
\usepackage[%
        colorlinks=true,urlcolor=blue,citecolor=blue,
        pdfpagelabels=true,hypertexnames=true,
        plainpages=false,naturalnames=true
        ]{hyperref}

\renewcommand{\figurename}{Figure}

\makeatletter
\def\maketitle{
\@author@finish
\title@column\titleblock@produce
\suppressfloats[t]}
\makeatother

\usepackage{xr}
\begin{document}
\title{Understanding the role of anharmonic phonons in diffusion of bcc metal}
\author{Seyyedfaridodin Fattahpour}
\author{Ali Davariashtiyani}
\author{Sara Kadkhodaei}
\email[]{To whom correspondence should be addressed; Email: sarakad@uic.edu}
\affiliation{Civil, Materials, and Environmental Engineering, University of Illinois at Chicago, 2095 Engineering Research Facility, 842 W. Taylor St., Chicago, IL 60607 USA}
\date{\today}

\begin{abstract}
Diffusion in high-temperature bcc phase of IIIB-IVB metals such as Zr, Ti, and their alloys is observed to be orders of magnitude higher than bcc metals of group VB-VIB, including Cr, Mo, and W. The underlying reason for this higher diffusion is still poorly understood. To explain this observation, we compare the first-principles-calculated parameters of monovancy-mediated diffusion between bcc Ti, Zr and dilute Zr-Sn alloys and bcc Cr, Mo, and W. Our results indicate that strongly anharmonic vibrations promote both the vacancy concentration and the diffusive jump rate in bcc IVB metals and can explain their markedly faster diffusion compared to bcc VIB metals. Additionally, we provide an efficient approach to calculate diffusive jump rates according to the transition state theory (TST). The use of standard harmonic TST is impractical in bcc IIIB/IVB metals due to the existence of ill-defined harmonic phonons, and most studies use classical or \textit{ab initio} molecular dynamics for direct simulation of diffusive jumps. Here, instead, we use a stochastically-sampled temperature-dependent phonon analysis within the transition state theory to study diffusive jumps without the need of any direct molecular dynamics simulations. We validate our first-principles diffusion coefficient predictions with available experimental measurements and explain the underlying reasons for promotion of diffusion in bcc IVB metals/alloys compared to bcc VIB metals.
\end{abstract}
\maketitle

\section{Introduction}\label{sec:introduction}
Diffusion in crystalline solids determines the kinetics of many diffusion-controlled phenomena, e.g., phase transformation, precipitate growth and coarsening, and corrosion, and understanding it is basic to predicting new materials with desired physical and mechanical properties. Diffusion in bcc solids is generally faster than close-packed phases of fcc and hcp due to their more open structure. However, within the bcc metals and their alloys, there are major differences in diffusion coefficient values. Specifically, self and solute diffusion is orders-of-magnitude faster in bcc phases of IIIB-IVB metals compared to VB-VIB metals~\cite{MURDOCK19641033,Peterson1978,herzig1987,herzigNeuhaus1987,Vogl1984,Vogl1989,PEART1962123,Petry1988,Herzig99,Nag2020}. For example, self-diffusion in bcc Ti is $10^5-10^7$ times higher than bcc Cr despite similar mass and common lattice structure~\cite{herzig1987}. The underlying reason for this strikingly faster diffusion is still not fully understood. Specifically, the effect of strong anharmonic phonons on diffusion is under-studied, considering that there is a fundamental difference between the nature of vibrations of bcc IIIB-IVB metals and VB-VIB metals. The bcc phase of the former group (IIIB-IVB metals and alloys) is only stable when it reaches high enough temperatures and experiences strongly anharmonic vibrations, i.e, it is dynamically stabilized~\cite{PETRY198956,PhysRevB.29.1575,RevModPhys.84.945,PhysRevB.95.064101,PhysRevB.61.11221,PhysRevLett.58.1769}, as opposed to the bcc phase in the latter group (VB-VIB metals) which is stable at all temperatures. This dynamical stabilization despite the lattice instability implies that the Hamiltonian anharmonicity is so strong that it creates multiple local minima around a high-symmetry maximum. The system is stabilized by hopping among these local minima at elevated temperatures, as shown in our earlier studies of thermodynamic stability of bcc Ti and ordered bcc NiTi compounds~\cite{PhysRevB.95.064101,KADKHODAEI2018296}. Aside from existing gaps in understanding the effect of phonon anharmonicity on diffusion, application of the harmonic transition state theory (TST) becomes impossible due to the existence of ill-defined harmonic phonons, i.e., phonons with imaginary frequencies or negative thermal energy, in dynamically stabilized phases~\cite{Eyring1935,VINEYARD1957121,Mantina2008}.   
%

Several studies have investigated the underlying reasons for the so-called anomalously fast diffusion in bcc IIIB-IVB metals and alloys, which we classify into three groups. The first group has explained the enhanced diffusion by mixed vacancy mechanisms, the appreciable contribution of divacancy or self-interstitial jumps to the vacancy mechanism especially near the melting point~\cite{MURDOCK19641033,Askill1965,Peart1967,Vogl1984,Willaime1990,Smirnov2020}. For example, Vogl et al. obtained two distinct jump frequencies for Co in bcc Zr from their quasi-elastic neutron scattering and speculated that jumps between interstitial sites might be the reason for the fast diffusion of Co~\cite{Vogl1984}. Recently, Smirnov has observed formation of self-interstitial defects in a classical molecular dynamics simulation based on a modified embedded atom method (MEAM) potential model for bcc Ti, showing that the self-interstitial jumps can contribute up to 10\% to the high-temperature diffusivity but become negligible at low temperature ranges~\cite{Smirnov2020}. Other EAM-potential MD studies consistently predicted the formation of self-interstitial in bcc Ti and Zr at high temperatures~\cite{Mendelev2010,Mendelev2016}. According to theses studies, self-interstitial jump contribution can explain the upward curvature of the Arrhenius plot for diffusion coefficient in bcc Ti and Zr at high temperature ranges, however, the markedly higher diffusivity compared to bcc VB-VIB, such as Cr or W, cannot be understood. Multiple isotope effect measurements and neutron scattering studies in bcc IIIB-IVB metals confirmed that the vacancy mechanism is the predominant diffusion mechanism~\cite{PEART1962123,Jackson1977,Manke1982,Petry1988,Vogl1989,petry1991}, precluding explanations based on major contributions from other intrinsic defects, even at temperatures close to the melting point~\cite{petry1991}. This lead to the second group of studies, which has related the anomalously fast and non-Arrhenius diffusion to soft phonon modes or to fluctuation of bcc IIIB-IVB metals between metastable $\omega$-phase embryos. Sanchez and de Fontaine proposed a model which correlates the diffusivity in bcc Zr to the formation of the metastable $\omega$ phase and uses the formation free energy of the $\omega$ embryo as the diffusion activation energy~\cite{Sanchez1975,SANCHEZ19781083,Sanchez1981}. Herzig and co-workers have explained the anomalously fast diffusion and strong non-Arrhenius behavior in bcc IIIB-IVB metals according to the characteristic soft phonon mode LA $\frac{2}{3}\left<111\right>$, and provided a semi-empirical relation between diffusive migration enthalpy and the square of the soft phonon frequency ~\cite{herzig1987,Kohler1987,KohlerHerzig1988}. Kadkhodaei and Davariashtiyani have illustrated that including the anharmonic phonon effects, obtained from \textit{ab initio} molecular dynamics, to describe the activation energy and the effective frequency for a vacancy-mediated diffusive jump in bcc Ti and Zr can successfully reproduce the experimentally reported anomalous high diffusion coefficients~\cite{Kadkhodaei2020}.
%
%
The third group of studies has related the enhanced diffusion to concerted motion of atoms or collective diffusion mechanisms~\cite{Sangiovanni2019,FRANSSON2020770}. Sangiovanni et al. observed a highly concerted string-like atomic motion in an \textit{ab initio} molecular dynamics simulation in bcc Ti at 1800 K~\cite{Sangiovanni2019}. They predicted diffusion coefficient of bcc Ti by accounting for both vacancy diffusivity and concerted atomic motion diffusivity obtained form MEAM-potential MD simulations, showing that the concerted motion diffusivity become non-negligible close to the melting point and can describe the non-Arrhenius curvature. Recently, MEAM-potential MD simulations in bcc and fcc metals have demonstrated that vacancy-interstitial pairs form via a string-like atomic motion along close-packed directions, which becomes particularly prevalent in bcc crystal structure, including Ti, Zr, Ta, and Nb~\cite{FRANSSON2020770}. According to these simulations, concentration of vacancy-interstitial pairs in bcc Zr, Ta, and Nb are in the same order, with bcc Ti showing exceptionally higher concentration and a strong deviation from the Arrhenius behavior at low temperature ranges. This study has linked the string-like atomic motion to vacancy-interstitial pair formation, both observed separately in previous MEAM MD simulations~\cite{Sangiovanni2019,Smirnov2020}. Despite these studies, a systematic comparison of diffusion parameters between ``normal'' bcc metals of VB-VIB and ``anomalous'' bcc metals of IIIB-IVB is still lacking. 


%

The contribution of this study is twofold: First, we elucidate the underlying reason for the so-called anomalously faster diffusion in bcc IVB metals/alloys with a focus on the role of strongly anharmonic vibrations. We present a detailed comparison of diffusion parameters between mechanically stable bcc VIB metals and mechanically unstable but dynamically stabilized bcc IVB metals, including first-principle-calculated activation energies and effective vibration frequencies along the diffusive jump direction. We explain the role of strongly anharmonic vibrations in promoting diffusive jumps. Second, we provide an efficient approach to calculate the diffusion coefficient of bcc IVB metals without the need of any direct simulation of diffusive jumps. We employ temperature-dependent phonon analysis within the TST to predict diffusive jump frequencies, where the temperature-dependent phonons are obtained based on a stochastic sampling of vibrations, eliminating any molecular dynamics simulation for sampling the vibrations or diffusive jumps. The content of this article is arranged in the following manner: In section~\ref{sec:method}, we describe the methods used for temperature-dependent phonon analysis and density functional theory (DFT) calculations to predict diffusion coefficient. In section~\ref{sec:results}, we illustrate the calculated diffusion parameters for bcc Cr, Mo, and W (as examples of mechanically stable or the so-called normal bcc metals) versus bcc Ti, Zr, and Zr-0.46at.\%Sn (as examples of dynamically stabilized or the so-called anomalous bcc metals). In section~\ref{sec:discussion}, we interpret the differences in diffusion parameters between normal and anomalous bcc systems in light of anharmonic lattice vibrations. Additionally, we discuss the findings of this study in comparison to existing studies in the literature.

\section{Method}\label{sec:method}
We obtain the macroscopic diffusion coefficient, $D$, of self-diffusion or solute-diffusion by a monovacancy mechanism in a bcc lattice according to the microscopic parameters in the following equation: 
\begin{equation}\label{eq:diffusivity}
D=C_v d^2 \Gamma
\end{equation}
where $d$ is the vacancy (or atom) jump distance, $C_v$ is the equilibrium concentration of vacancy or vacancy-solute pair, and $\Gamma$ is the successful vacancy jump rate~\cite{Mantina2008}. Vacancy jump distance in bcc is equal to the nearest neighbor distance or $\frac{\sqrt{3}}{2}a_0$, where $a_0$ is the lattice constant. The vacancy or vacancy-solute pair concentration at temperature $T$ is given by $C_v = \exp(\frac{\Delta S_f}{k_B})\exp(-\frac{\Delta H_f}{k_BT})$, where $\Delta H_f$ and $\Delta S_f$ are the formation enthalpy and entropy of vacancy or vacancy-solute pair, respectively, and $k_B$ is the Boltzmann constant. The vacancy jump rate $\Gamma$ is obtained from the migration enthalpy $\Delta H_m$ and the effective vibration frequency along the migration path $\nu^*$ by  $\Gamma = \nu^* \exp(\frac{-\Delta H_m}{k_BT})$. According to the TST, the effective vibration frequency, $\nu^*$, is the ratio of the product of normal vibration frequencies (or harmonic phonon frequencies) of the initial state of atomic migration, $\nu_i$ to that of the non-imaginary normal frequencies of the transition state, $\nu^{\prime}_j$, i.e., $\nu^* = \frac{\prod^{3N-3}_{i=1} \nu_i}{ \prod^{3N-4}_{j=1} \nu^{'}_j}$~\cite{VINEYARD1957121}. 

We calculate the microscopic parameters of diffusion, including $\Delta H_f$, $\Delta S_f$, $\Delta H_m$, and $\nu^*$, in different ways for the normal and anomalous bcc systems: For normal bcc metals, including Cr, Mo and W, we employ the standard approach~\cite{Mantina2008} based on density functional theory (DFT) total energy calculations, the climbing image nudged elastic band (c-NEB) method for locating the transition state~\cite{Henkelman2000}, and the frozen phonon method for harmonic phonon analysis~\cite{phonopy}. For anomalous bcc metals, we provide a new approach that uses a stochastic sampling of the canonical or NVT ensemble to describe vacancy formation enthalpy and entropy, vacancy migration enthalpy, and effective vibration frequency. Details of each approach are explained in the following.
\begin{figure}[!h]\includegraphics[width=1.0\textwidth]{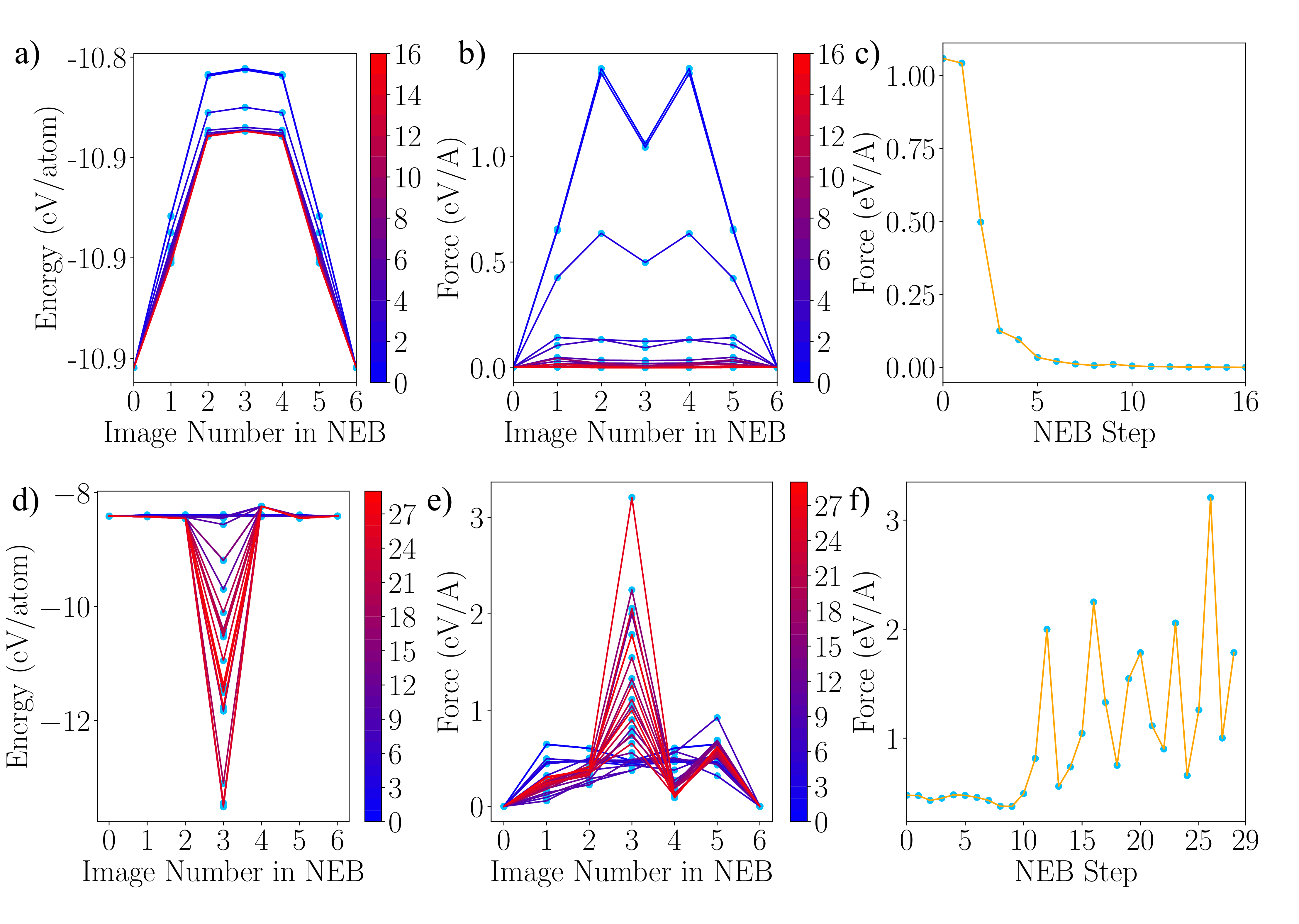}\hfill
\caption{{The evolution of the NEB to optimize the initial energy profile to the final minimum energy path for bcc Mo versus bcc Zr}. The initial band consists of 5 images interpolated along the $\frac{1}{2}[111]$ nearest neighbor vacancy jump (or diffusive jump) direction. a,d) The energy evolution on each image throughout the NEB optimization for bcc Mo and bcc Zr, respectively. b,e) The force evolution on each image for bcc Mo and bcc Zr, respectively. c,f) The maximum force on the climbing image at each NEB iteration for bcc Mo and bcc Zr, respectively.} %
\label{fig:NEB}
\end{figure}

For Cr, Mo, and W, the equilibrium lattice constant, $a_0$, is calculated by finding the minimum DFT energy for five different volumes incremented by 2\% lattice constant expansion. The vacancy formation enthalpy is calculated as the difference of DFT total energies of the defected and bulk systems, $E(N-1)$ and $E(N)$, according to $\Delta H_v = E(N-1) - \frac{N-1}{N}E(N)$, for a system of $N$ atoms. The vacancy formation entropy is calculated similarly according to $\Delta S_v = S^{\text{vib}}(N-1) - \frac{N-1}{N}S^{\text{vib}}(N)$. The vibration entropy, $S^{\text{vib}}$, is calculated from to the harmonic phonon density of states according to the Supplementary Equation~\ref{eq:free energy of vibration}. The harmonic phonon density of states are calculated from the temperature-independent force-constant using the frozen phonon approach as implemented in the Phonopy package~\cite{phonopy}. We use the climbing image NEB method based on DFT forces to obtain the minimum energy path and the saddle point (or the transition state of diffusion)~\cite{Henkelman2000}. Figure~\ref{fig:NEB}(a-c) shows the NEB optimization for bcc Mo, for example. Supplementary Figure~\ref{fig:MEP2} shows the NEB optimization for other bcc metals. Subsequent to the transition sate optimization, we compute the migration enthalpy, $\Delta H_m$, as the difference between the DFT total energies of the initial atomic configuration and the activated transition state, $\Delta H_m = E^{\text{activated}}(N-1) - E^{\text{initial}}(N-1)$.
The effective vibration frequency $\nu^*$ is obtained by calculating the normal vibration modes for the initial and the saddle point atomic configurations.
\begin{figure}[!h]\includegraphics[width=1.0\textwidth]{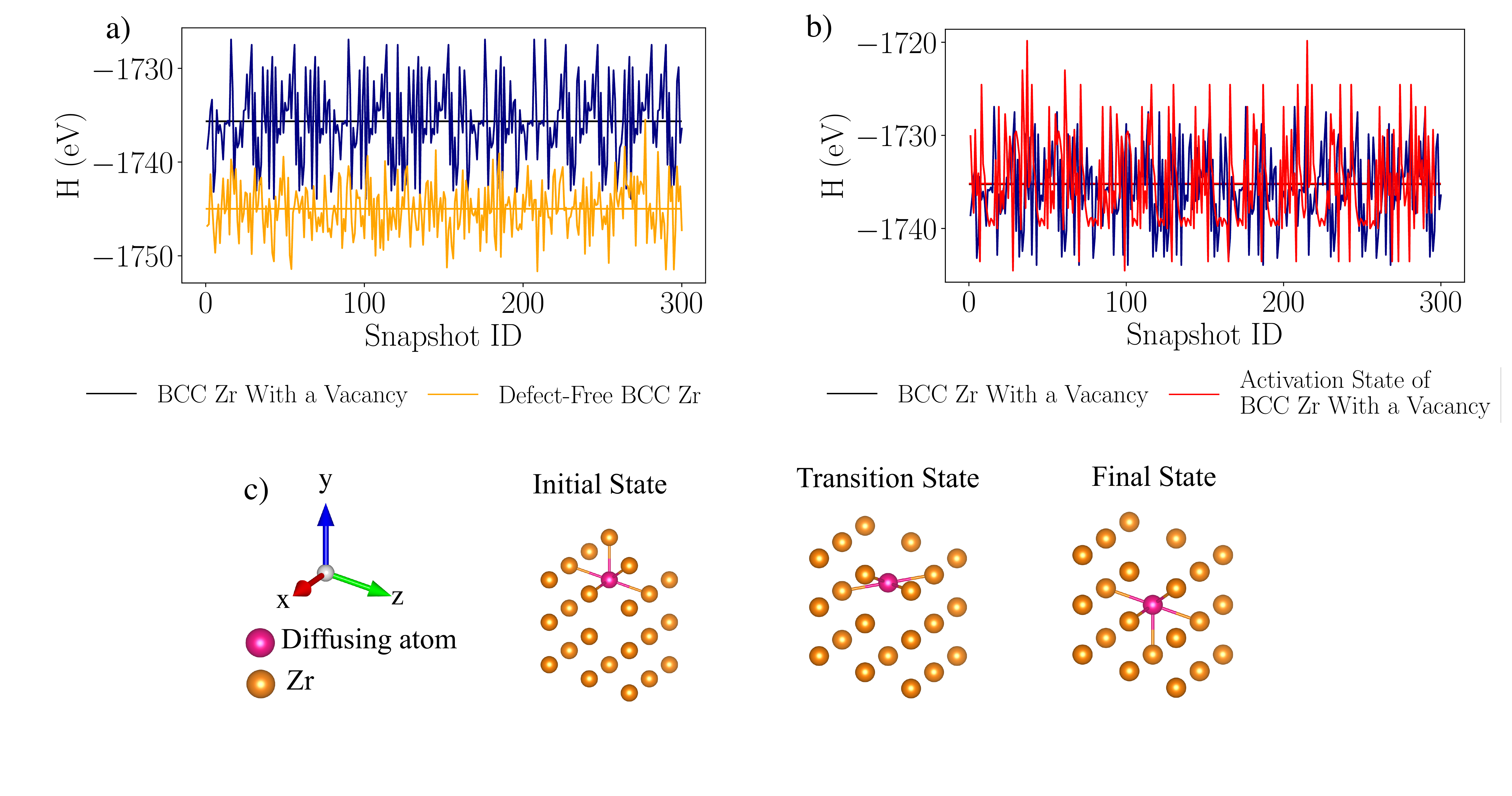}\hfill
\caption{Stochastic sampling of the canonical ensemble for dynamically stabilized bcc systems. a) Enthalpy at 1400 K for all the stochastic snapshots of the defect-free bcc Zr (blue) and defected bcc Zr with a monovacancy (orange). b) Enthalpy at 1400 K for all the stochastic snapshots of the defected bcc Zr with a monovacancy (blue) and the activated defected bcc Zr with the diffusing atom halfway between neighboring bcc lattice sites (orange). The horizontal lines represent averages over all the snapshots. c) Static atomic configurations, i.e., atoms sitting at the lattice sites without any momentum, around the vacancy for the initial, transition (or activated), and final states of diffusion. Diffusing Zr atom is shown by a different color. The static transition state is approximated by an atomic configuration where the diffusing atom sits between two vacant neighboring lattice sites.}
\centering
\label{fig:stochastic}
\end{figure}

\begin{figure}[!h]\includegraphics[width=1.0\textwidth]{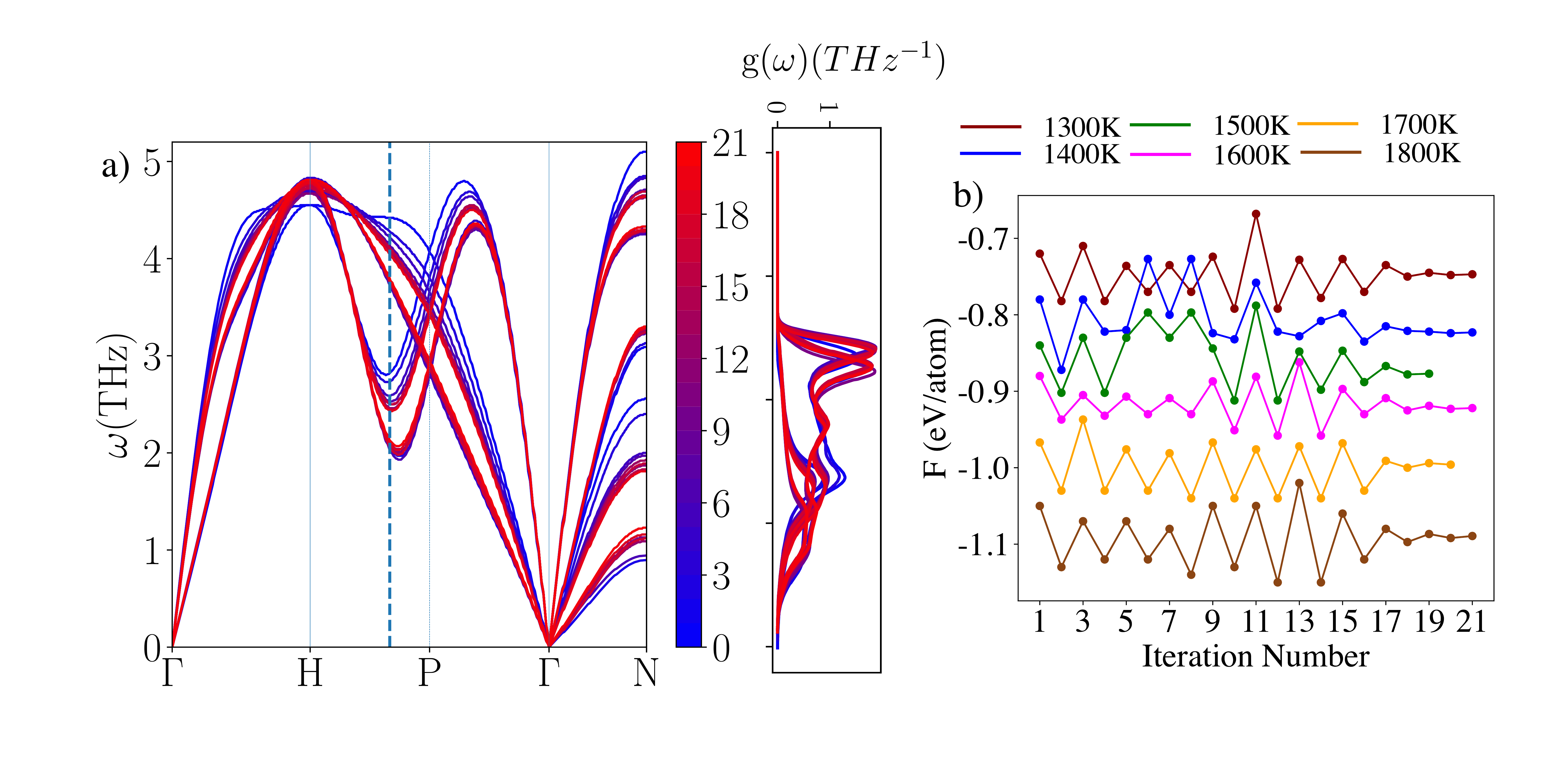} \hfill
\caption{Iterative temperature-dependent phonon analysis for the defect-free bcc Zr. a) Phonon dispersion and density of states evolution during the s-TDEP iterations at 1400 K. b) Phonon free energy evolution during the s-TDEP iterations for bcc Zr at different temperatures. }
\centering
\label{fig:phonon_dispersions}
\end{figure}
For dynamically stabilized bcc Ti, Zr, and Zr-0.46at.\%Sn, we provide an approach that can effectively include the temperature-dependent anharmonic vibrations into the diffusion calculation. Due to the mechanical instability of the bcc lattice in these systems, the formation or migration enthalpy cannot be approximated by the total energy difference of bulk and defected or activated supercells, considering that the DFT energy corresponds to the free energy at absolute zero temperature, where these system do not even exist. An additional difficulty arising from the lattice instability is the divergence of saddle point search schemes based on DFT forces. As compared in Figure~\ref{fig:NEB}(a,b) and (d,e), the NEB method can optimize an initial energy profile along the $\frac{1}{2}[111]$ nearest neighbor vacancy jump direction to the final minimum energy path and therefore can locate the transition state in bcc Mo, while it fails to locate the transition state in bcc Zr based on DFT forces. Figure~\ref{fig:NEB}(f) shows that the maximum force on the climbing image diverges for bcc Zr (see Supplementary Figure~\ref{fig:MEP2} for bcc Ti), whereas Figure~\ref{fig:NEB}(c) shows the convergence of the maximum force on the climbing image in bcc Mo (see Supplementary Figure~\ref{fig:MEP2} for bcc Cr and W). 
To overcome these challenge, we generate three different canonical (or NVT) ensembles for each anomalous bcc metal. One for the defect-free or bulk supercell, one for the defected supercell (including a monovacany or a vacancy-solute pair), and one for the activated transition state. 
The canonical ensembles are generated using the Maxwell-Boltzmann statistics according to the stochastic temperature-dependent effective potential (or s-TDEP) method~\cite{Hellman2011,Hellman2013,Hellman2013Oct,Shulumba2017,Yang2020}. The atomic positions and velocities are generated using the harmonic normal mode transformation, as detailed in Supplementary Note 1 and Supplementary Equations~\ref{eq:canonical configurations} and ~\ref{eq:thermalAmpl}, resulting in uncorrelated excited states (or snapshots) for each ensemble. Unlike the bulk and equilibrium defected states, the activated transition state is a semi-equilibrium state, for which the canonical ensemble is obtained by generating multiple atomic snapshots with stochastic positions and velocities, except for the diffusing atom which has a fixed position at halfway along the $\frac{1}{2}\left[111\right]$ diffusive jump direction but has a stochastic velocity. This construct assumes that the saddle point on an effective temperature-dependent energy surface coincides with the atomic configuration with diffusing atom at $\frac{1}{4}\left[111\right]$ lattice point (see Figure~\ref{fig:stochastic}(c)). We use DFT to calculate atomic forces, total energy, and pressure for each stochastic snapshot in the ensembles. For the canonical ensemble at 1400 K, we calculate the Helmholtz free energy from the stochastic snapshots (as described below) for 6 different volumes incremented by $0.02a_0$ and select the volume with the lowest free energy as the equilibrium volume for the canonical ensemble (same method used in Ref.~\cite{Yang2020}). We use the same volume for supercells at other temperatures.  
The enthalpy values are obtained by averaging the total energy (sum of the ion-electron and the kinetic energies), $\langle U\rangle$, and the pressure-volume term, $\langle p\rangle V$, over the snapshots for each ensemble according to $H=\langle U\rangle +\langle p\rangle V $, as shown in Figure~\ref{fig:stochastic}. The vacancy formation and migration enthalpy values are then obtained according to $\Delta H_v = H(N-1) - \frac{N-1}{N}H(N)$ and $\Delta H_m = H^{\text{activated}}(N-1) - H^{\text{initial}}(N-1)$, respectively. We confirmed that the average total energy obtained from the stochastic snapshots agrees well with the average total energy obtained from a canonical \textit{ab initio} molecular dynamics simulation. More details are provided in the Supplementary Figure~\ref{fig:MD_vs_TDEP}. 
For all the bcc metals, we incorporate the intrinsic surface correction terms to the vacancy formation and migration enthalpy values as shown in Supplementary Table ~\ref{tab:energy_corrections}. These correction terms are calculated based on the method proposed in Ref.~\cite{Mattsson2002}, and compensate for the underestimation of DFT GGA in calculating  the intrinsic surface energy that is formed around a vacancy. 
\begin{figure}[!h]\includegraphics[width=1.0\textwidth]{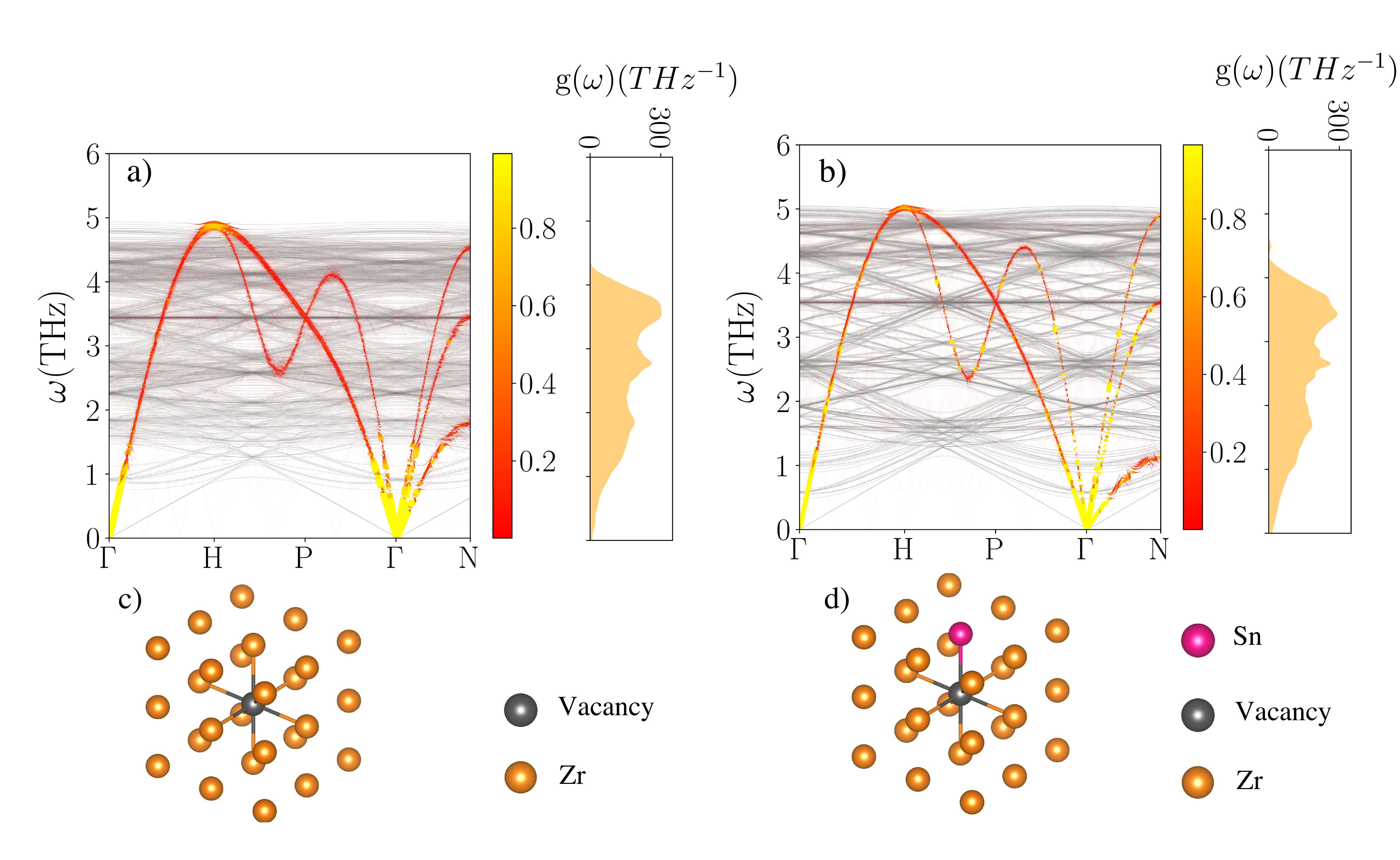} \hfill
\caption{The temperature-dependent phonon dispersion and density of states for a) the defected bcc Zr at 1400K and b) the defected bcc Zr-0.46at.\%Sn at 1400 K. The folded dispersion is shown by gray and the unfolded dispersion is shown by a color map that indicates the spectral function of the unfolding. The atomic configuration of c) the defected bcc Zr around the isolated vacancy and d) the defected bcc Zr-0.46at.\%Sn around the vacancy-solute pair. }
\centering
\label{fig:unfolded_phonon_dispersions}
\end{figure}

For bcc Ti, Zr, and Zr-Sn systems, phonons at elevated temperatures are calculated from the effective (or converged) temperature-dependent second-order force-constant according to the TDEP method~\cite{Hellman2011,Hellman2013,Hellman2013Oct}. The displacements and forces on stochastic atomic snapshots are recorded, and the force constant is obtained via a least-squares fitting of an effective harmonic potential model of Supplementary Equation~\ref{eq:hamiltonian} to the displacement-force data set. The force constant is converged through a self-consistent process with respect to the number of snapshots and iterations. At each temperature, 50 stochastic snapshots are generated initially using a model force-constant that emulates the Debye temperature (as explained in Ref.~\cite{Shulumba2017}). The calculated force constant at each iteration is then used to generate a new set of stochastic snapshots for the next iteration (see Supplementary Equations~\ref{eq:canonical configurations} and ~\ref{eq:thermalAmpl}). The number of snapshots is increased as the iteration continues to ensure the convergence of the phonon free energy, with the last iteration consisting of 300 stochastic snapshots. Throughout the s-TDEP iterations, the phonon free energy is calculated using the Supplementary Equation~\ref{eq:free energy of vibration}. Convergence is achieved when the phonon free energy difference between consecutive iterations falls below 3 meV per atom. Figure~\ref{fig:phonon_dispersions} shows the evolution of the phonon dispersion, phonon density of states, and phonon free energy through the s-TDEP iterations for the defect-free bcc Zr.  
%
%
The vacancy formation entropy, $\Delta S_f$, is calculated based on the difference of the vibration entropy for the defected and bulk systems, each calculated from the temperature-dependent phonon density of states (e.g., see Figure~\ref{fig:unfolded_phonon_dispersions} and Supplementary Figure~\ref{fig:Zr_215}) according to the Supplementary Equation~\ref{eq:free energy of vibration}. The effective prefactor frequency along the diffusive jump direction, $\nu^*$, is approximated as the temperature-dependent phonon frequency at $\frac{2}{3}$ along the $\text{L}\left<\xi \xi \xi\right>$ or $\Gamma$-H-P branch for the defected system. The atomic distortion for this phonon mode coincides with the vacancy jump direction for the bcc phase. Figure~\ref{fig:unfolded_phonon_dispersions} shows the temperature-dependent phonon dispersion and density of states calculated at 1400 K for defected bcc Zr and Zr-0.46at.\%Sn. For defected systems, $\nu^*$ or phonon frequency at $\frac{2}{3}\text{L}\left<111\right>$ is obtained from the unfolded phonon dispersion. For phonon band unfolding, we use the method of Ref.~\cite{Ikeda2017} by imposing the bcc unit cell symmetry path of '$\Gamma$', 'H', 'P', '$\Gamma$', 'N' to the defected supercell (see Figure~\ref{fig:unfolded_phonon_dispersions}). 

Details of the DFT and NEB calculations are as the following: For Cr, Mo, and W, a $5\times5\times5$ supercell of conventional bcc unit cell with 250 atoms is used for the bulk system. For bcc Ti, Zr, and Zr-Sn, a $6\times6\times6$ supercell containing 216 atoms is used for the bulk state, and the defected supercell consists of 215 atoms. We use the projector-augmented-wave method (PAW) as implemented in the highly efficient Vienna \textit{ab initio} simulation package (VASP)~\cite{PAW,vasp1,vasp2,vasp3}. For Cr, Mo, and W, we use a Monkhorst-Pack k-point mesh of $3\times3\times3$ and an energy cutoff of 350 and 320 and 300, respectively,  within the PBE exchange-correlation functional. For Zr, Ti, and Zr-Sn, a $1\times1\times1$ Monkhorst-Pack k-point grid, and a plane-wave energy cutoff of 350 eV within the PBE exchange-correlation functional~\cite{Ernzerhof1999}. We use the spring constant of 5.0 eV/$Å^2$ for all the NEB calculations and convergence is assumed when the maximum atomic force on all the images falls below 0.01 eV/$\AA$. 
\section{Results}\label{sec:results}
\begin{figure}[!h]
\includegraphics[width=1\textwidth]{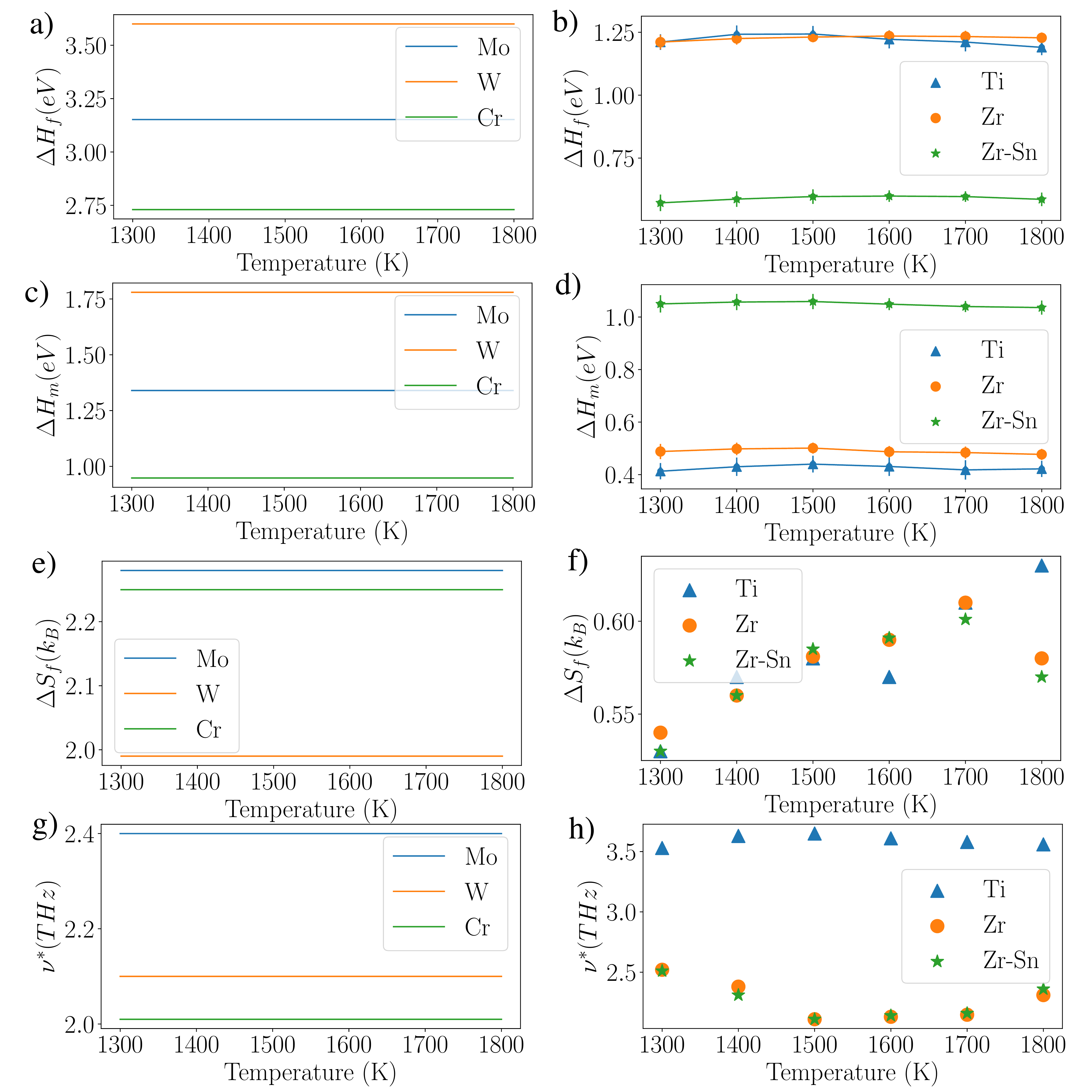}
\caption{Microscopic diffusion parameters for mechanically stable bcc metals (left column) and dynamically stabilized bcc systems (right column). a, b) Vacancy formation enthalpy, $\Delta H_f$. c,d) Vacancy migration enthalpy, $\Delta H_m$. e, f) Vacancy formation entropy, $\Delta S_f$. g, i) Effective prefactor frequency, $\nu^*$. Error bars indicate the standard deviation of the stochastic snapshots at each temperature (for each NVT ensemble).}
\centering
\label{fig:Diffusion_parameter}
\end{figure}

We calculate the individual microscopic diffusion parameters, namely the vacancy formation enthalpy, $\Delta H_f$, vacancy migration enthalpy, $\Delta H_m$, vacancy formation entropy, $\Delta S_f$, and effective prefactor frequency, $\nu^*$, for the mechanically stable versus dynamically stabilized bcc systems as explain in section~\ref{sec:method}. As shown in Figure~\ref{fig:Diffusion_parameter}, diffusion parameters for the mechanically stable bcc metals (Cr, Mo, and W) are temperature-independent because the formation and migration enthalpies are calculated from DFT total energies of bulk and defected atomic configurations and formation entropy and effective prefactor frequency are calculated from the harmonic phonon density of states and harmonic TST. On the other hand, diffusion parameters for dynamically stabilized bcc Ti, Zr, and Zr-0.46at.\%Sn show temperature-dependence as expected when calculated from stochastically sampled canonical ensembles and temperature-dependent phonon analysis. 

\begin{figure}[!h]\includegraphics[width=1.0\textwidth]{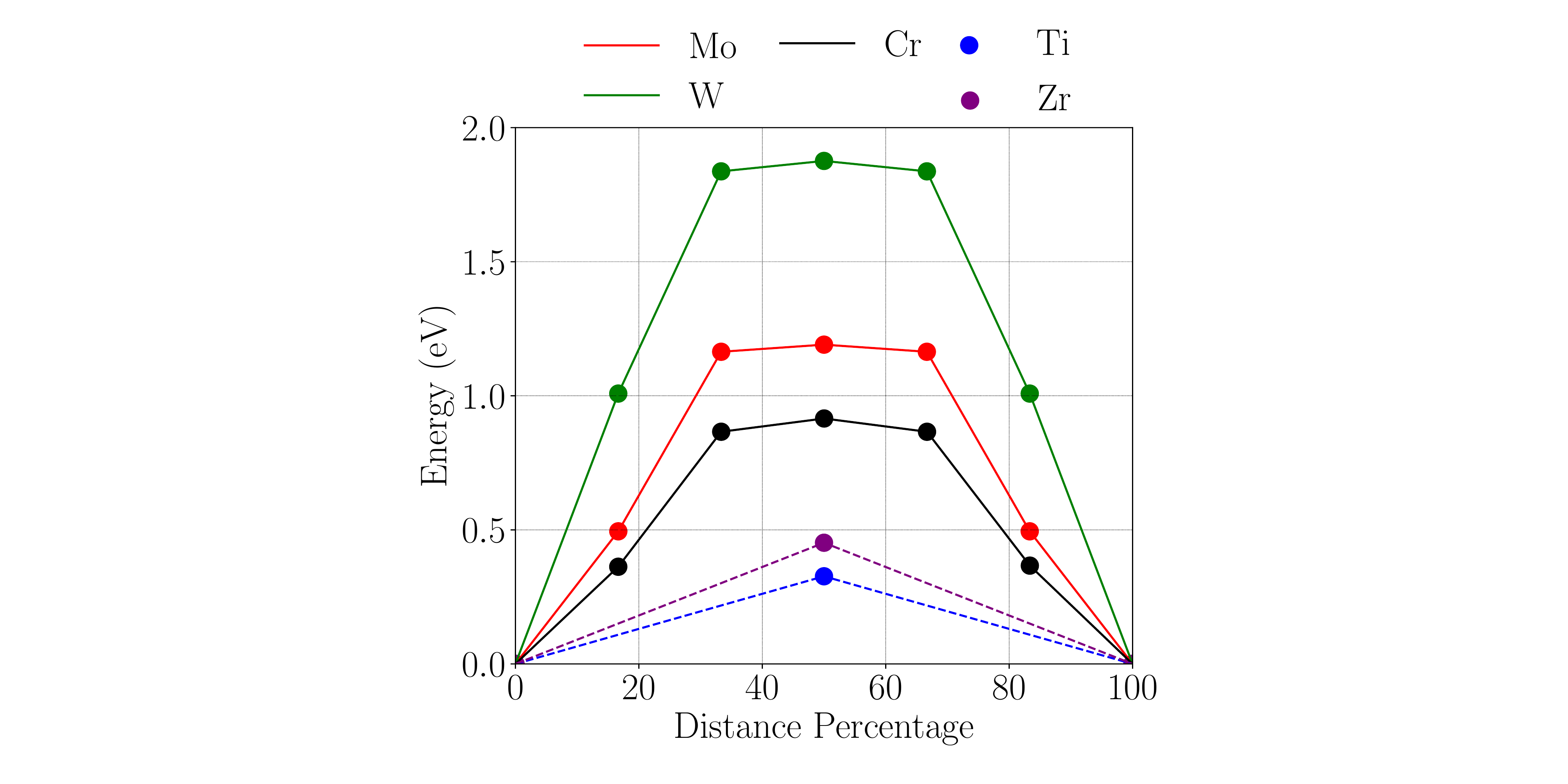}\hfill
\caption{DFT energy versus distance percentage along the vacancy diffusion pathway. For Mo, W and Cr the energy profile is optimized using the c-NEB. For Ti and Zr, the represented energy barrier is the DFT energy between the static atomic configurations of the initial and the transition state, approximated by fixing the diffusing atom at the $\frac{1}{4}[111]$ position. The calculated migration barrier for bcc Ti and Zr are calculated from average enthalpies over canonical stochastic snapshots instead of static configurations (see Figure~\ref{fig:stochastic}) .}
\centering
\label{fig:MEP}
\end{figure}

The formation and migration enthalpies for dynamically stabilized bcc phases are lower than their mechanically stable counterparts, as illustrated in Figure~\ref{fig:Diffusion_parameter}(a) and (b) and (c) and (d). For example, $\Delta H_f$ (or $\Delta H_m$) is 2.73 eV (or 0.948 eV) for bcc Cr compared to 1.242 eV (or 0.43 eV) at 1400 K for bcc Ti with a similar atomic size or it is 3.152 eV (or 1.34 eV) for bcc Mo compared to 1.225 eV (or 0.498 eV) at 1400 K for bcc Zr (read from Table~\ref{tab:Diffusion_parameters}). Among mechanically stable bcc metals formation and migration enthalpies increase by atomic size, lowest for Cr and largest for W (see Figure~\ref{fig:Diffusion_parameter}(a,c) and Figure~\ref{fig:MEP}). However, in between dynamically stabilized systems, the atomic size has a reduced significance. As shown in Figure~\ref{fig:Diffusion_parameter} (b,d), the formation enthalpy values for bcc Ti and Zr are almost the same and the migration enthalpy for Ti is only slightly higher than Zr. However, $\Delta H_f$ for Zr-Sn is largely decreased compared to Ti and Zr while $\Delta H_m$ exhibits a large increase. This is because of the large binding energy between tin solute and vacancy (1.225 eV - 0.56 eV = 0.66 eV at 1400 K, read from Table~\ref{tab:Diffusion_parameters}). The attractive nature of the binding decreases the formation energy and subsequently increases the vacancy-solute concentration (as reported in Table~\ref{tab:Diffusion_parameters}). On the other hand, the attractive binding energy resists the solute-vacancy exchange or migration and increases $\Delta H_m$ compared to the monovacancy migration in Zr or Ti. The relative lower activation enthalpies (formation + migration) of bcc Ti, Zr, and Zr-Sn is due to the dynamically stabilized nature of these phases. In fact, they reside on high-energy regions of the potential energy surface as opposed to a deep well (or low energy local minimum) in mechanically stable bcc metals. This implies a flat energy profile along the migration pathway (see Figure~\ref{fig:MEP}) which relatively reduces $\Delta H_m$ and a shallow well of residence which relatively reduces $\Delta H_f$. The shallow nature of the residing well implies the existence of soft bonds or phonons with lower frequencies for dynamically stabilized bcc phases, leading to $\Delta S_f$ of about 3 to 4 times lower than mechanically stable bcc systems with the more pronounced vacancy-induced softening effect (see Figure~\ref{fig:Diffusion_parameter} (e,f)). $\nu^*$ for dynamically stabilized bcc metals is only slightly higher than their mechanically stable counterparts (see Figure~\ref{fig:Diffusion_parameter} (g,h) and Table~\ref{tab:Diffusion_parameters}).
%
%

In Table~\ref{tab:Diffusion_parameters}, we compare the calculated diffusion parameters with available values in the literature. For bcc Cr, Mo, and W, our predictions of $a_0$, $\Delta H_f$, $\Delta H_m$, and $\Delta S_f$ are in perfect agreement with other DFT predictions~\cite{Mattsson2009,Yasuhiro2014,MEDASANI201596}, with the only exception of $\Delta H_f$ for bcc Cr being lower than the DFT study in Ref.~\cite{MEDASANI201596}.
For bcc Ti and Zr, DFT data are few due to computational difficulties associated with the lattice instability. Therefore, we compare our predictions with available data that are directly measured or calculated from other measurements (more details are provided in Table~\ref{tab:Diffusion_parameters}). Our $\Delta H_f$ for bcc Ti and Zr are relatively lower than the formation enthalpies evaluated from the nonlinear increase in specific heat via modulation measurements in Ref.~\cite{KRAFTMAKHER199879} but show the same trend between Ti and Zr. Our $\Delta H_m$ values for bcc Ti and Zr are slightly higher than the values evaluated from measured phonon density of states in Ref.~\cite{Schober1992} but $\Delta H_m$ for bcc Zr closely match the prediction from a quench molecular dynamics study~\cite{Willaime1990}. Our $\Delta S_f$ for bcc Ti and Zr are lower than the evaluations from Born-von Karman fits to the measured phonon dispersion curves~\cite{Schober1992} (see Table~\ref{tab:Diffusion_parameters}), however, they closely match predictions from 0 K calculations of quenched molecular dynamic snapshots of Ref.~\cite{Willaime1990}, as reported in Table~\ref{tab:Diffusion_parameters}. Our $\Delta S_f$ for bcc Cr and W, on the other hand, are in better agreement with values of Ref.~\cite{Schober1992}. The evaluated parameters of Ref.~\cite{Schober1992} only provides estimates and should not be considered as quantitative, especially in cases of strong anharmonictiy of phonons (see section 6.3. in Ref.~\cite{Schober1992}). 
\begin{table}[ht]
\caption{Diffusion Parameters of bcc Mo, W, Cr, Zr, Ti, and Zr-Sn. The following superscripts are used:  $**$: Calculated value at 1400K in this work,  $a$: DFT simulation with AM05 functional~\cite{Mattsson2009}, $b$: DFT Simulation with PBE functional~\cite{Yasuhiro2014}, $c$: DFT Simulation with PBE functional~\cite{MEDASANI201596},  $d$: Experimental value from modulation measurements of specific heat~\cite{KRAFTMAKHER199879},  $f$: From static lattice Green function which is directly calculated from experimental phonon density of states \cite{Schober1992},  $g$: From force-constants fitted to the measured phonon dispersion curves \cite{Schober1992}, and  $h$: Values at 0 K from quenched molecular dynamics~\cite{Willaime1990}.} 
\begin{center}
\resizebox{\textwidth}{!}{%
\begin{tabular}{cc|c|c|c|c|c|c|}
\cline{3-8}
\multicolumn{1}{l}{}                                                     & \multicolumn{1}{l|}{} & Mo                    & W                     & Cr                    & Ti                    & Zr                    & Zr-Sn(0.46at\%)    \\  
\hline
\multicolumn{1}{|c|}{\multirow{1}{*}{$a_0$ ($\AA$)}} 
                                                                                &  This Work     & 3.135                 & 3.185                 & 2.84                  & 3.28$^{**}$         &3.59$^{**}$                  & 3.59$^{**}$                  \\ \cline{2-8} 
                                                                                \multicolumn{1}{|c|}{}                                                          & Reference      & 3.134$^a$                 & 3.186$^b$                 & 2.84$^c$                  & 3.25$^c$                    & -                     & -                       \\ \hline
\multicolumn{1}{|c|}{\multirow{1}{*}{$\Delta H_f$ (eV)}}                      &  This Work     & 3.152                 & 3.6                   & 2.73                  & 1.242$^{**}$     & 1.225$^{**}$                 & 0.578$^{**}$              \\ \cline{2-8} 
\multicolumn{1}{|c|}{}                                                          & Reference      & 3.1$^a$-3.07$^c$  & 3.51$^c$  &              3.05$^c$               & 1.55$^d$                     & 1.75$^d$-1.53$^h$                     & -                    \\ \hline
\multicolumn{1}{|c|}{\multirow{1}{*}{$\Delta S_f$ ($k_B$)}}            & This Work      & 2.28      & 1.99      & 2.25      & 0.57$^{**}$          & 0.56$^{**}$      & 0.56$^{**}$                  \\ \cline{2-8} 
\multicolumn{1}{|c|}{}                                                          & Reference      & 2.3$^a$                   & 1.8$^g$                   & 1.8$^g$                    & 2.4$^g$(1293K)                   & 2.58$^g$(1483K) - 0.5$^h$                 & -                      \\ \hline
\multicolumn{1}{|c|}{\multirow{1}{*}{$\Delta H_m$ (eV)}}                      & This Work      & 1.34                  & 1.78                  & 0.948                & 0.43$^{**}$         & 0.498$^{**}$                 & 1.057$^{**}$     \\ \cline{2-8} 
\multicolumn{1}{|c|}{}                                                          & Reference      & 1.3$^a$                   & 1.78$^b$                  & 0.95$^c$                 & 0.31$^f$(1293K)                  & 0.324$^f$(1483K)-0.32$^h$                  & -                     \\ \hline
\multicolumn{1}{|c|}{\multirow{1}{*}{$\nu^*$  (THz)}}                                 & This Work      & 2.4                   & 2.1                   & 2.01                 & 3.63$^{**}$                  & 2.38$^{**}$                 & 2.31$^{**}$            \\ \cline{2-8} 
 \hline
\multicolumn{1}{|c|}{\multirow{1}{*}{$C_v$}}                                   & This Work      &6.90E-11$^{**}$               & 8.02E-13$^{**}$             & 1.409E-09$^{**}$              &  5.97E-05$^{**}$                   & 6.81E-05$^{**}$              & 1.34E-2$^{**}$                   \\ \cline{2-8} 
\hline
\multicolumn{1}{|c|}{\multirow{1}{*}{$\Gamma$(Hz)}}                  & This Work      & 3.6E+07$^{**}$           & 8.21E+05$^{**}$            & 1.15E+09$^{**}$            & 1.02E+11$^{**}$          & 3.5E+10$^{**}$           & 3.61E+08$^{**}$                      \\ \cline{2-8} 
\hline
\end{tabular}
}
\label{tab:Diffusion_parameters}
\end{center}
\end{table}

\begin{figure}[!h]\includegraphics[width=1\textwidth]{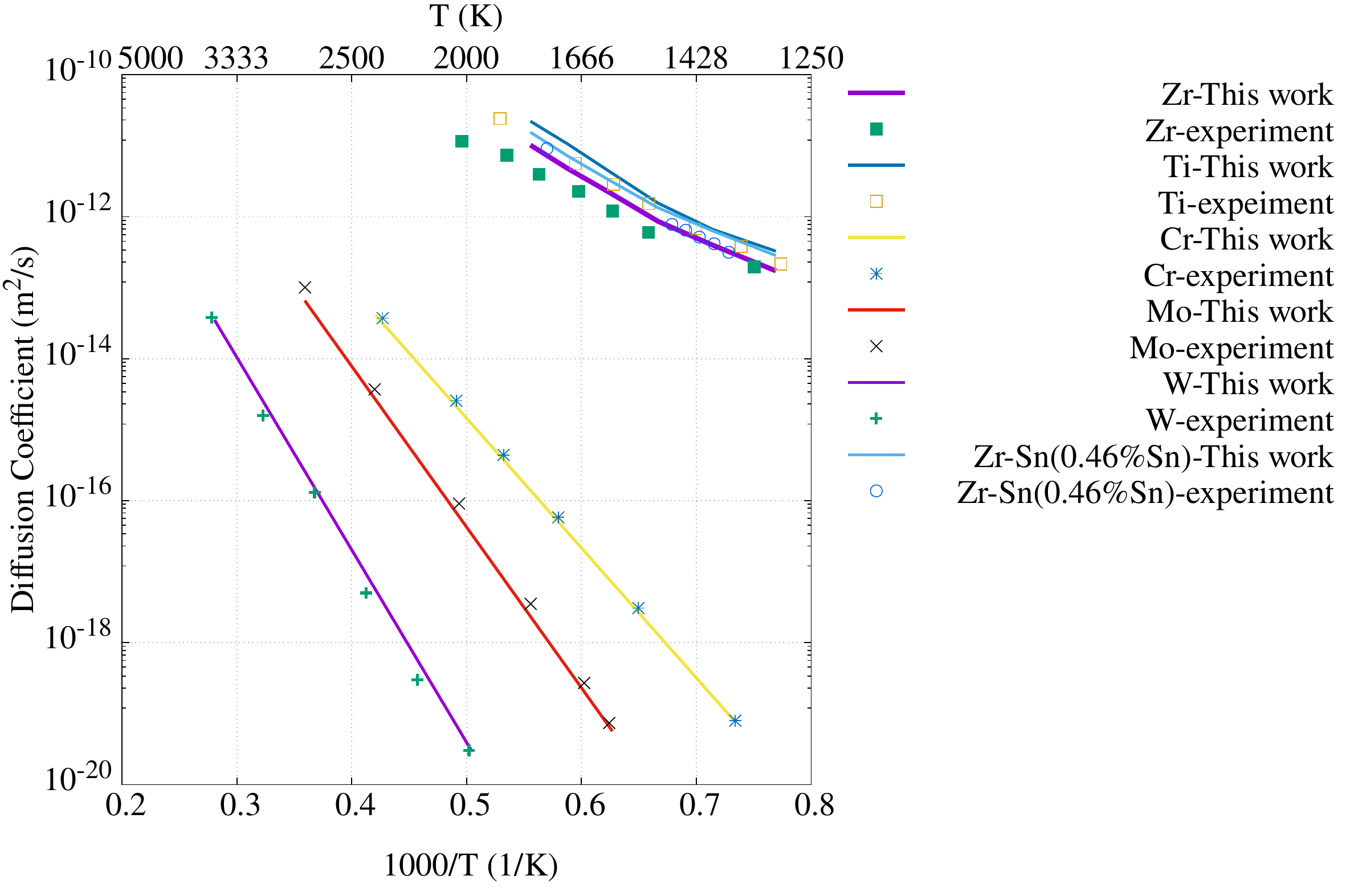}
\caption{Self and solute diffusion coefficients versus temperature in bcc IVB and VIB metals. The predictions in this work are from first-principles calculations, compared against experimental measurements of Ref.~\cite{herzig1987} for pure systems and Ref.~\cite{zhu2019measurement} for the Zr-Sn alloy. }
\centering
\label{fig:Diffusion}
\end{figure}

In addition to microscopic diffusion parameters, we report the vacancy (or vacancy-tin pair) concentration and the successful vacancy jump rate, $C_v$ and $\Gamma$, as defined in Section~\ref{sec:method}. $C_v$ and $\Gamma$ for dynamically stabilized bcc systems are both orders of magnitude higher than their mechanically stable counterparts (see reported values at 1400 K in Table~\ref{tab:Diffusion_parameters}). For example, $C_v$ at 1400 K is 6.8E-05 for bcc Zr compared to 6.90E-11 for bcc Mo (almost $10^6$ times higher) or 5.97E-05 for bcc Ti compared to 1.409E-09 for bcc Cr (more than 40,000 times higher). Similarly, $\Gamma$ at 1400 K is 3.5E+10 Hz for bcc Zr compared to 3.6E+07 Hz for bcc Mo (around 850 times higher) or it is 1.02E+11 Hz for bcc Ti compared to 1.15E+09 Hz for bcc Cr (around 100 times higher). The markedly higher vacancy concentration and successful diffusive jump rate both contribute to the strikingly higher diffusivity in dynamically stabilized bcc phases compared to the mechanically stable ones (see Figure~\ref{fig:Diffusion}). Among mechanically stable bcc phases, $C_v$ exhibits a strong dependence on the atomic size, similar to $\Delta H_f$ as expected, with highest vacancy concentration in Cr and lowest in W. Unlike mechanically stable bcc metals, $C_v$ is less sensitive to atomic size in dynamically stabilized bcc systems, similar to the weak dependence of $\Delta H_f$ on atomic size in these phases. 
%

Figure~\ref{fig:Diffusion} shows the calculated diffusion coefficients from the diffusion parameters according to Equation~\ref{eq:diffusivity} for all bcc systems compared against available experimental values. Dynamically stabilized bcc systems, Ti, Zr, and Zr-Sn, show around $10^5-10^6$ times higher diffusion coefficients compared to mechanically stable bcc metals, Cr, Mo, and W. The markedly higher diffusivity is more pronounced at lower temperatures, as expected based on the Arrhenius $D$-$T$ relation. Another remarkable difference is that diffusion coefficients of bcc Ti, Zr, and Zr-Sn only span over a narrow range ($10^{-11}-10^{-13}$ m$^2$/s) while diffusivities of bcc Cr, Mo, and W span over a wider range ($10^{-14}-10^{-20}$ m$^2$/s). This is because of the lower downward slope of the $D$-$T$ curves in the former group arising from lower activation enthalpies. 
Among bcc Ti, Zr, and Zr-Sn, $D$ values are very close, unlike bcc Cr, Mo, and W, where diffusivities are distinctly different with a strong dependence on the size of diffusing atom, lowest for W and highest for Cr. Among anomalous bcc systems, the order of self or solute diffusion from high to low is Ti, Zr-Sn, Zr. The slight higher diffusivity of Ti compared to Zr can be attributed to higher vacancy jump frequency (e.g., 3.34 times higher in Ti compared to Zr at 1400 K) while vacancy concentration of Ti and Zr are almost the same (compare values in Table~\ref{tab:Diffusion_parameters}). The slight higher diffusivity of Zr-Sn compared to Zr, on the other hand, arises from the trade-off between the increased vacancy concentration in Zr-Sn ($\approx$200 times higher than Zr at 1400 K) and the reduced vacancy jump frequency ($\approx$ 0.01 times lower than Zr at 1400 K). This is because of the large attractive binding between tin solute and vacancy, which significantly favors vacancy-solute pair formation but it increases the migration enthalpy.

\begin{figure}[!h]
\includegraphics[width=0.5\textwidth]{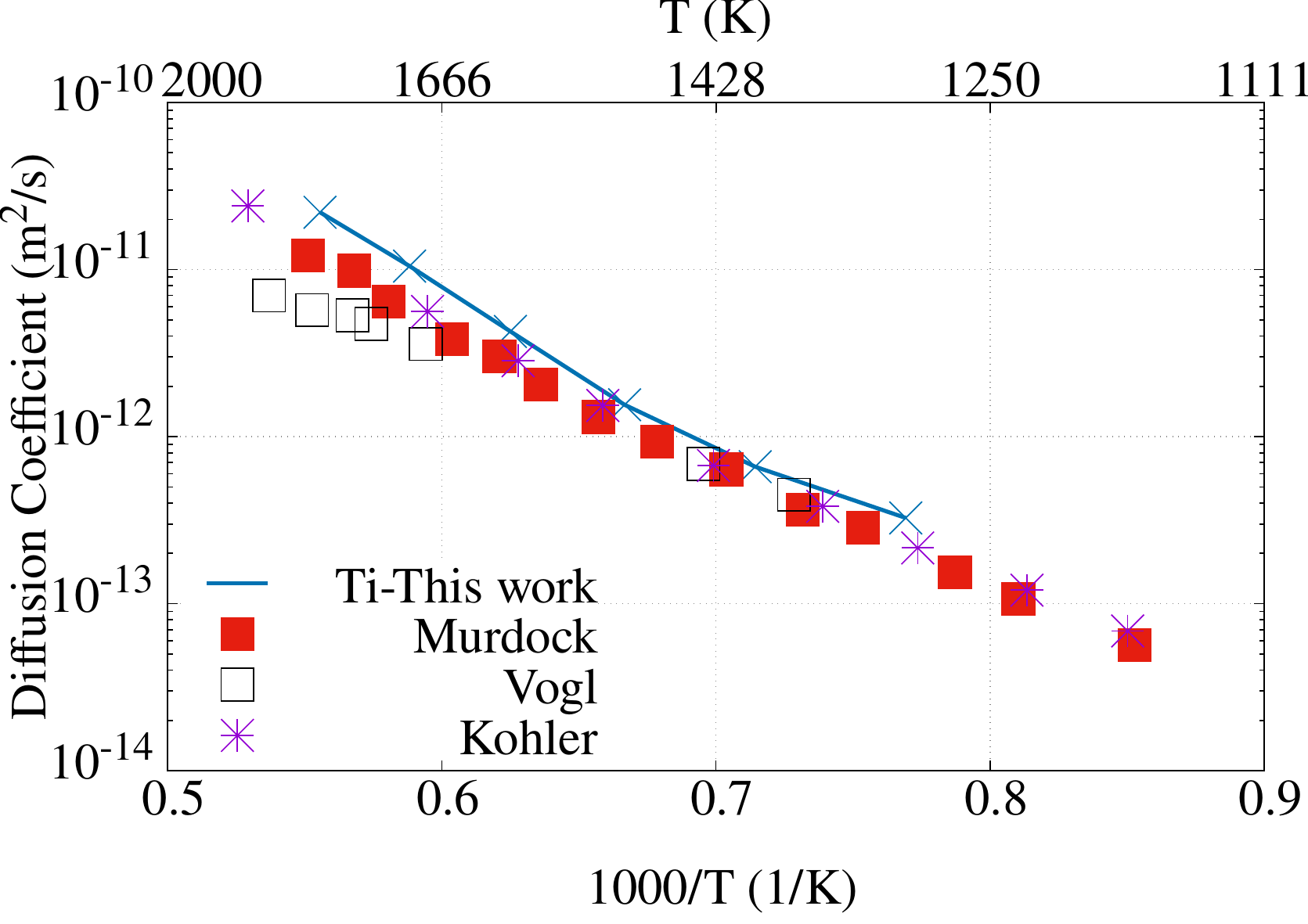}
\includegraphics[width=0.5\textwidth]{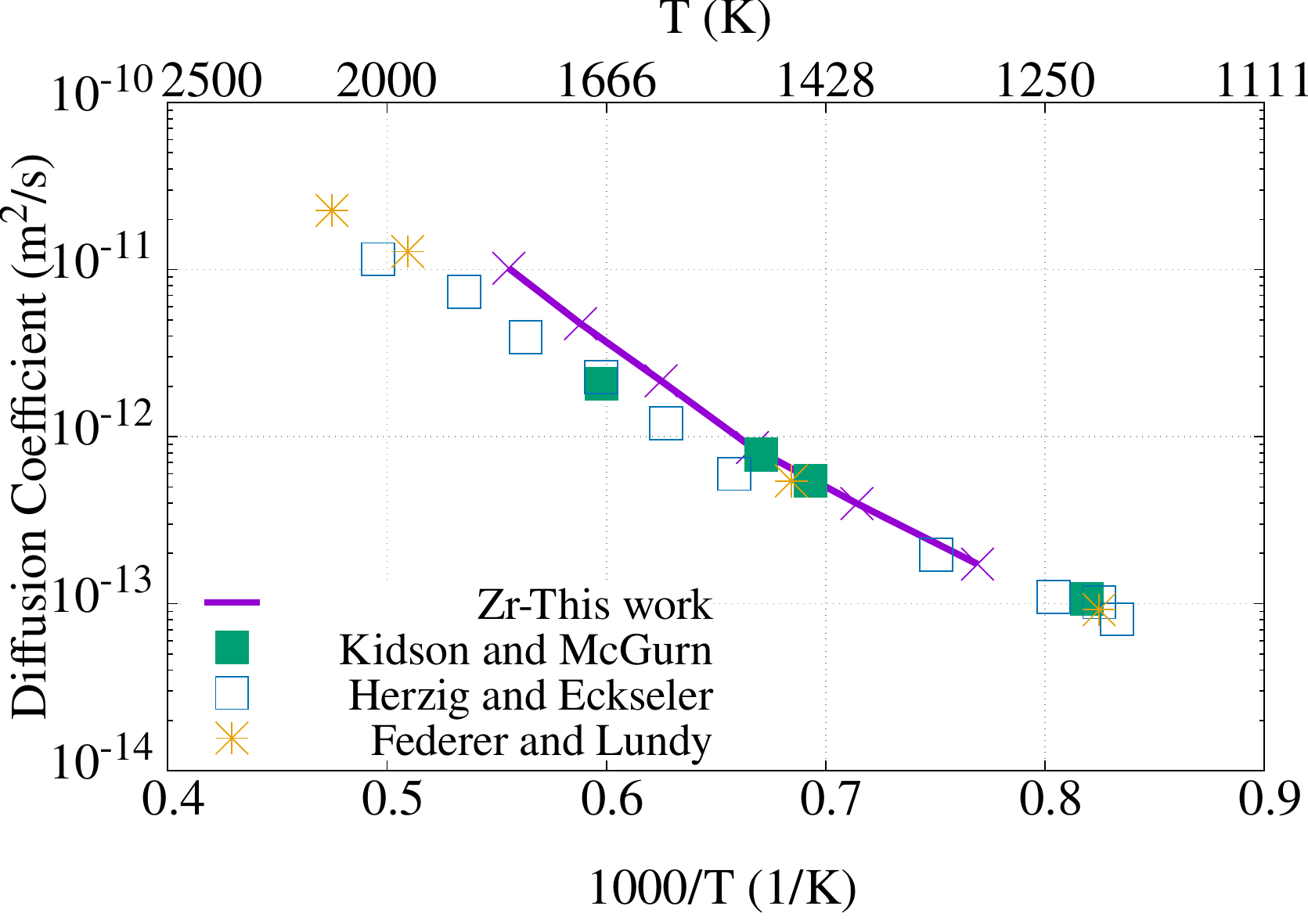}

\caption{Self-diffusion in bcc Ti (top) and bcc Zr (bottom) versus temperature. The references to experimental results are Murdock et al.\cite{MURDOCK19641033}, Vogl et al.\cite{Vogl1989}, Kohler et al.\cite{Kohler1987}, Kidson and Mc Gurn\cite{KidsonandMcGurn}, Herzig and Eckseler\cite{herzig1979anomalous}, and Federer and Lundy\cite{federer1963diffusion}}
\centering
\label{fig:Diffusion_Ti_Zr}
\end{figure}
As shown in Figure~\ref{fig:Diffusion}, our first-principles predictions of diffusion coefficient are in good agreement with experimental measurements~\cite{herzig1987,zhu2019measurement}. For self-diffusion in bcc Ti and Zr, a more detailed comparison of our diffusivity predictions with several experimental measurements~\cite{MURDOCK19641033,Vogl1989,Kohler1987,KidsonandMcGurn,herzig1979anomalous,federer1963diffusion} is shown in Figure~\ref{fig:Diffusion_Ti_Zr}. Our calculated diffusion coefficients agree well with all measured values and show a slight curvature in the Arrhenius plots. In Supplementary Figure~\ref{fig:non_linearity}, we evaluate the dependence of the curvature of Arrhenius plots on the temperature-dependence of different diffusion parameters in bcc Ti and Zr. 
%

%
\section{Discussion}\label{sec:discussion}
Our first-principles calculations indicate that the anomalously higher diffusivity in dynamically stabilized bcc phases stems from both the increased vacancy concentration and successful diffusive (or vacancy) jump rate. Both increases are a direct result of the strongly anharmonic nature of vibrations in these phases: 1) Monovacany formation is significantly promoted in these phases because of the nature of free energy surface which corresponds to a shallow well. In fact, the bcc lattice instability is so strong that it creates multiple local minima on the energy surface around the bcc structure associated with local lattice distortions, and the system is effectively residing on a high-energy shallow well by hopping among the local structural distortions. 2) Monovacany diffusive jump rate is significantly promoted due to the coincidental softening of restoring forces along the $\frac{1}{2}\left[111\right]$ diffusive jumps, a direct result of strong anharmonicity of phonons (or phonon softening) along this direction (compare the phonon dispersion curves for bcc Zr, Zr-Sn, and Ti with those for bcc Cr, Mo, W in Figure~\ref{fig:unfolded_phonon_dispersions} and Supplementary Figure~\ref{fig:allDispersion}). 

We incorporate the strongly anharmonic vibration effects in the description of diffusion parameters, which enable us to predict the individual parameters of vacancy concentration and vacancy jump rate. This elucidates the role of strong phonon anharmonicity in anomalously fast diffusion in terms of two separate effects: 1) an effective shallow well of residence and 2) directional lattice softening along [111], both arising from strong lattice anharmonicity. Interestingly, these two effects are also the underlying reason for the hetero-phase fluctuations in anomalous bcc systems, i.e., the vibration-induced fluctuations between bcc and $\omega$, which is studies extensively in the literature~\cite{SANCHEZ19781083,hickman1969}. The earlier model by Sanchez and de Fontaine has effectively intertwined these two effects by relating the hetero-phase fluctuation to the anomalous diffusion. They assumed that the formation free energy of the $\omega$ phase is the diffusive activation barrier~\cite{Sanchez1975} based on the fact that the softening in L $\frac{2}{3}\left<111\right>$ phonon mode is associated with the low-frequency large-amplitude vibration that results in hetero-phase fluctuations in the same direction as diffusive hops in the bcc lattice. 
%

The good agreement of available experimental diffusion coefficients with our first-principles predictions based on the monovacancy jump mechanism supports the dominant role of monovacancy diffusion in bcc metals, consistent with earlier incoherent quasielastic neutron scattering experiments~\cite{petry1991}. Our calculations suggest that the contribution of other diffusion mechanisms to macroscopic diffusivity is secondary but can become important especially at high temperatures. Based on the presented calculations, we speculate that the secondary diffusion mechanisms are necessary for a more accurate prediction of the observed non-Arrhenius curvature in anomalous bcc systems. The temperature-dependence of activation enthalpy and the prefactor frequency only result in a small curvature in the Arrhenius plots. Secondary mechanisms of diffusion are extensively discussed in the literature, e.g., crowdion-like mechanism~\cite{Willaime1990}, interstitialcy mechansim~\cite{Smirnov2020}, or string-like collective mechanism~\cite{Sangiovanni2019}. 

\begin{acknowledgments}
We would like to thank Dr. O. Hellman for kindly sharing the TDEP package with us. This work was supported by the US National Science Foundation Award Number DMR-1954621. We used the Extreme Science and Engineering Discovery Environment (XSEDE) through allocation TG-MAT200013, which is supported by National Science Foundation grant number ACI-1548562 resources~\cite{xsede}.
\end{acknowledgments}

\bibliographystyle{unsrt}

\clearpage

\setcounter{equation}{0}
\setcounter{figure}{0}
\setcounter{table}{0}
\setcounter{section}{0}
\renewcommand{\thetable}{\arabic{table}}
\renewcommand{\thefigure}{\arabic{figure}}
\renewcommand*\thesection{Supplementary Note \arabic{section}}

\renewcommand*\thesubsection{\thesection.\arabic{subsection}}

\renewcommand{\figurename}{Supplementary Figure}
\renewcommand{\tablename}{Supplementary Table}

\title{Supplementary Information for \textit{Understanding the role of anharmonic phonons in diffusion of bcc metal}}
\author{Seyyedfaridodin Fattahpour}
\author{Ali Davariashtiyani}
\author{Sara Kadkhodaei}
\email[]{To whom correspondence should be addressed; Email: sarakad@uic.edu}
\affiliation{Civil, Materials, and Environmental Engineering, University of Illinois at Chicago, 2095 Engineering Research Facility, 842 W. Taylor St., Chicago, IL 60607 USA}
\date{\today}

\maketitle

\section{Temperature-dependent Phonon Analysis}\label{sec:tdep}

We use the stochastic temperature-dependent effective potential (s-TDEP) method for studying lattice dynamics at elevated temperatures~\cite{Shulumba2017}. We employ s-TDEP for stochastic sampling of uncorrelated canonical micro-states of the simulation cell. This stochastic sampling yields a force-displacement data set from which an effective force constant that best describe the following harmonic potential model is obtained via a least-square fitting process.

\begin{equation}\label{eq:hamiltonian}
\hat{U}= U_0+
\frac{1}{2!}\sum_{ij} \sum_{\alpha\beta}\Phi_{ij}^{\alpha\beta}
u_i^\alpha u_j^\beta
\end{equation}

Here, $u_i$ is the displacement of atom $i$, $\alpha\beta$ are Cartesian components, $\Phi$ is the second order effective force constant, and $U_0$ is the reference energy. Using the stochastic approach in Ref.~\cite{Shulumba2017}, the supercell with $N$ atoms of mass $m_i$ is thermally populated with a set of displacements $\{u_i\}$ and velocities $\{\dot{u}_i\}$ given by

\begin{equation}\label{eq:canonical configurations}
\begin{split}
&u_i = \sum_{s=1}^{3N}
\epsilon_{is} \langle A_{is} \rangle \sqrt{-2\ln \xi_1}\sin 2\pi\xi_2 \\
&\dot{u}_i = \sum_{s=1}^{3N}
\omega_s \epsilon_{is} \langle A_{is} \rangle \sqrt{-2\ln \xi_1}\cos 2\pi\xi_2
\end{split}
\end{equation}
where $\xi_1$ and $\xi_2$ are uniformly distributed numbers between 0 and 1 and $\left<A_{is}\right>$ is the thermal amplitude of the normal mode $s$ with eigenvector $\epsilon_{is}$ and frequency $\omega_s$, given by

\begin{equation}\label{eq:thermalAmpl}
\langle A_{is} \rangle
\approx
\frac{1}{\omega_s}\sqrt{\frac{k_BT}{m_i}}.
\end{equation}

Throughout an iterative process, a stochastic set of force-displacement is generated using Equations~\ref{eq:canonical configurations} and ~\ref{eq:thermalAmpl} and is used in a least-square fitting of Equation~\ref{eq:hamiltonian} to identify the effective harmonic force constant $\Phi$. To generate the initial set of force-displacement in the absence of a fitted $\Phi$, we use a pair-wise potential model whose phonons zero-point energy reproduces the Debye temperature, as proposed in Ref.~\cite{Shulumba2017}. The self-consistent iterative process is continued until the phonon free energy obtained from the effective force constant is converged. The Helmholtz free energy for the canonical ensemble is $F = U_0 + F_{\text{vib}}$, where $F_{\text{vib}}$ is the phonon free energy given by
\begin{equation}\label{eq:free energy of vibration}
F_{\textrm{vib}}= -T S_{\textrm{vib}} = \int_{0}^{\infty} g(\omega) \left\{ k_B T \ln\left[ 1-\exp\left(-\frac{\hbar\omega}{k_BT}\right)\right]+\frac{\hbar\omega}{2}\right\} d\omega,
\end{equation} 
where $g(\omega)$ is the phonon density of states, calculated from the phonons in the first Brillouin zone:
\begin{equation}\label{eq:dos}
g(\omega) = \int_{\mathrm{BZ}} \delta( \omega - \omega_{\mathbf{q}s}) d\mathbf{q}.
\end{equation}

\begin{table}[H]
\caption{Correction terms added to the vacancy formation and migration enthalpy to compensate for the underestimation of DFT GGA calculation in the presence of an intrinsic surface formed around a vacancy. The intrinsic surface correction term per area as a function of bulk valence electron density of PBE exchange-correlation function is obtained from Ref.\cite{Kadkhodaei2020} according to the original method of Ref.~\cite{Mattsson2002}. }
\setlength{\tabcolsep}{10pt} 
\renewcommand{\arraystretch}{1.5} 
\resizebox{\textwidth}{!}{%
\begin{tabular}{|c|c|c|c|c|c|c|c|}
\hline
{\textbf{Element}} &
  {\textbf{Nelect}} &
  {\textbf{\begin{tabular}[c]{@{}c@{}}Number of\\  atoms \\ in the bcc \\ unit cell\end{tabular}}} &
  {\textbf{Lattice constant (A)}} &
  {\textbf{Bulk Density}} &
  {\textbf{\begin{tabular}[c]{@{}c@{}}Intrinsic\\ surface\\ correction\\ term per\\ area (eV/$\AA^2$)\end{tabular}}} &
  {\textbf{\begin{tabular}[c]{@{}c@{}}Formation \\ Correction \\ (eV)\end{tabular}}} &
  {\textbf{\begin{tabular}[c]{@{}c@{}}Barrier\\ Correction\\ (eV)\end{tabular}}} \\ \hline
{\textbf{Cr}}                   & 6 & 2 & 2.84  & 0.523  & 0.022 & 0.297 & 0.065 \\ \hline
{\textbf{Mo}}                   & 6 & 2 & 3.135 & 0.389 & 0.0191 & 0.303 & 0.068 \\ \hline
{\textbf{W}}                    & 6 & 2 & 3.185 & 0.371  & 0.0172 & 0.305 & 0.069 \\ \hline
{\textbf{Zr}}                   & 4 & 2 & 3.59 & 0.172   & 0.0094 & 0.187 & 0.049 \\ \hline
{\textbf{Ti}}                   & 4 & 2 & 3.28 & 0.226   & 0.0121 & 0.182 & 0.053 \\ \hline
\end{tabular}%
}
\label{tab:energy_corrections}
\end{table}

\begin{figure}[H]\includegraphics[width=1.0\textwidth]{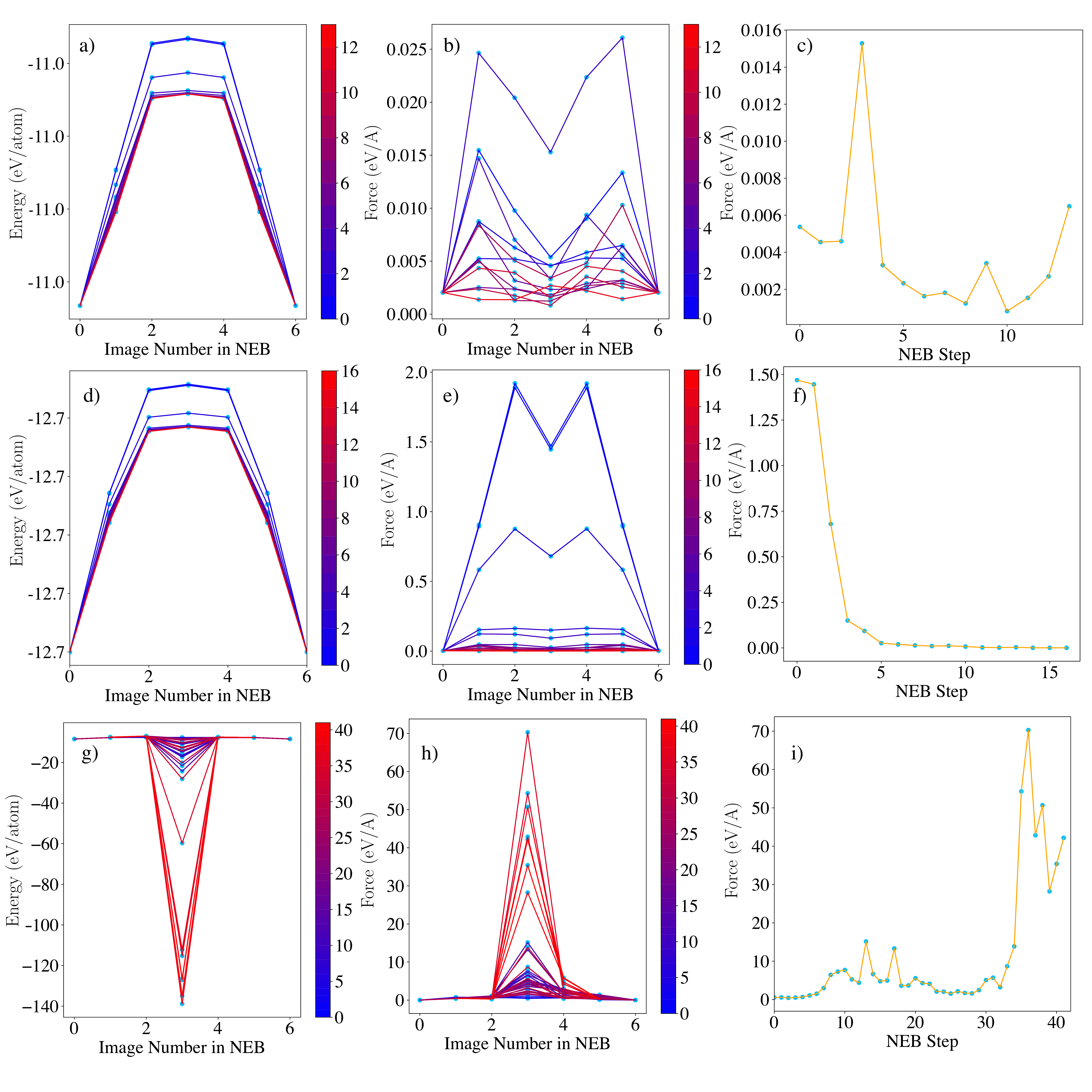}
\caption{
The evolution of the c-NEB iterations from the initial energy profile to the final minimum energy path. The initial band consists of 5 images interpolated along the $\frac{1}{2}[111]$ nearest neighbor vacancy jump (or diffusive jump) direction. The evolution of DFT energy profile (left column), DFT force profile (middle column), and the maximum force on the climbing image or image number 3 (right column) for a,b,c) bcc Cr, d,e,f) bcc W, g,h,i) bcc Ti. c-NEB converges for mechanically stable bcc Cr and bcc W but diverges for bcc Ti with lattice instability.}
\label{fig:MEP2}
\end{figure}

\begin{figure}[H]\includegraphics[width=\textwidth]{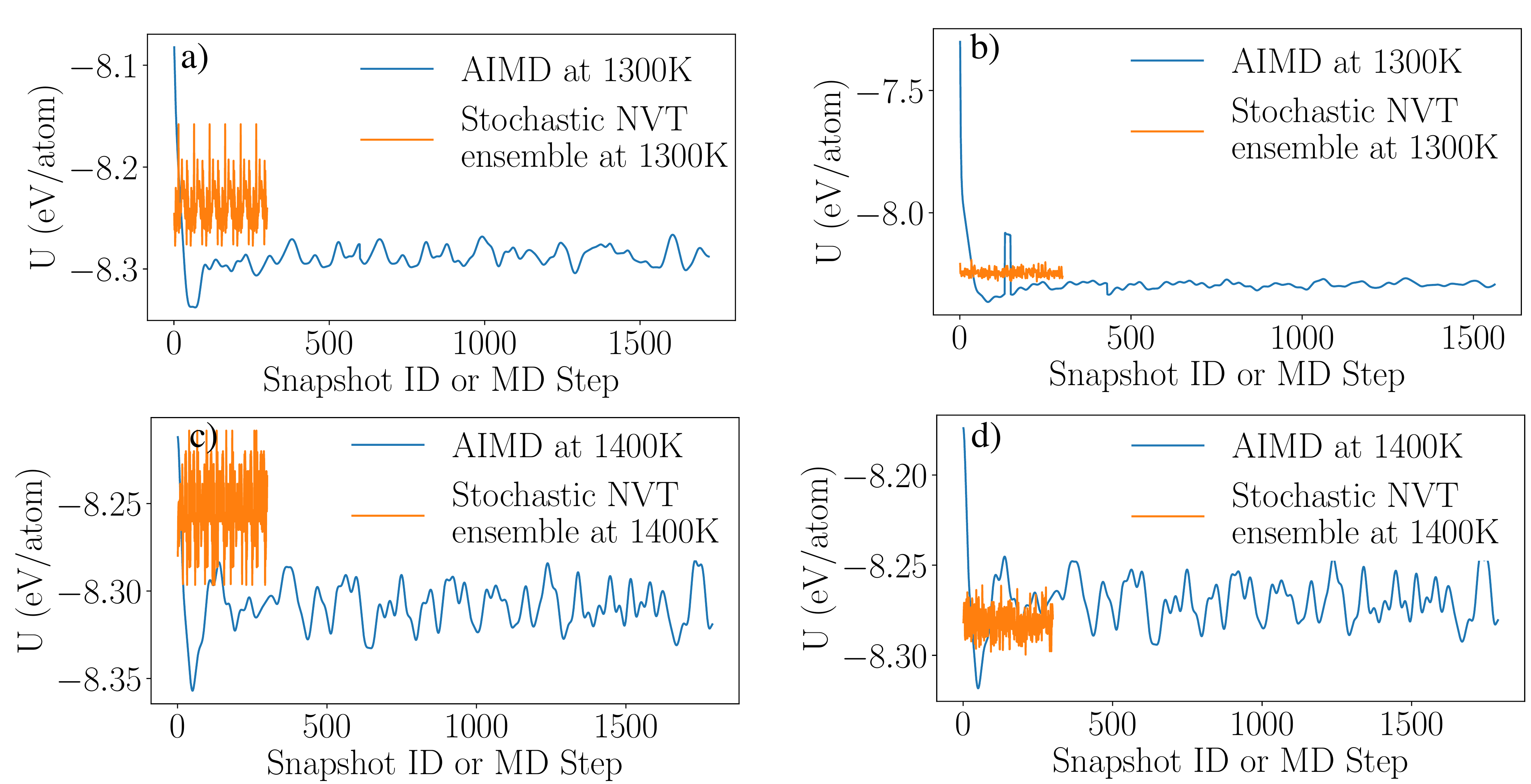}\vspace{3mm}
\caption{Comparison of the total energy $U$ (ion-electron + kinetic) obtained from stochastic canonical sampling versus NVT \textit{ab initio} molecular dynamics (AIMD) for bcc Zr. $U$ is presented versus stochastic snapshot identification number or AIMD step for a) the defected supercell with a monovacany at 1300 K, b) the defect-free bcc supercell at 1300 K, c) the defected supercell with a monovacany at 1400 K, and d) the defect-free bcc supercell at 1400 K. The average $U$ from AIMD and stochastic sampling perfectly agrees for the bulk system. $U$ of the defected system is slightly lower for AIMD. This is likely because of local atomic relaxation around the vacancy in AIMD simulations that is not accounted for in the stochastic sampling. The effect of this small deviation on formation enthalpy is negligible.}
\label{fig:MD_vs_TDEP}
\end{figure}

\begin{figure}[H]\includegraphics[width=\textwidth]{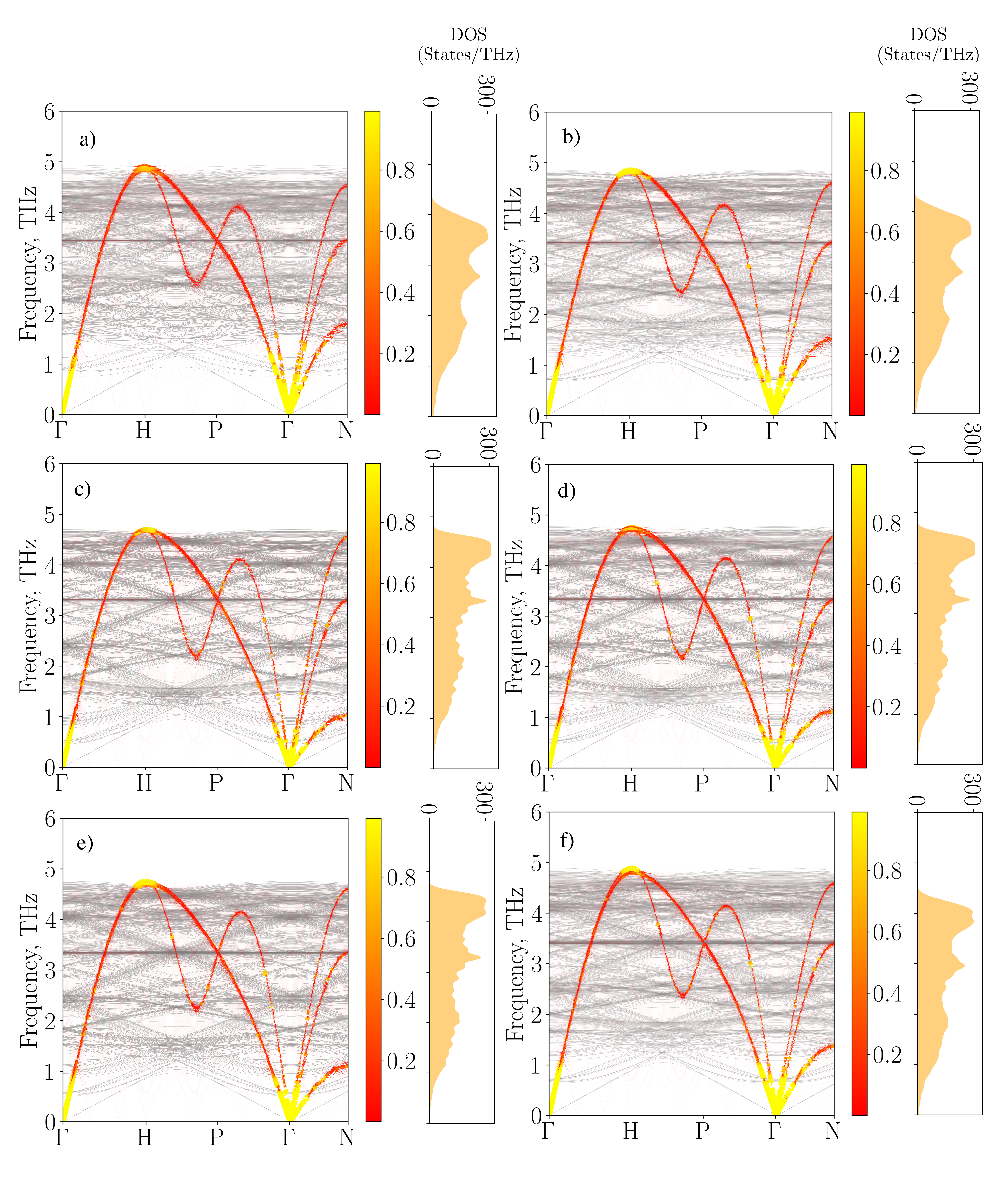}
\caption{The temperature-dependent phonon dispersion and density of states for defected bcc Zr, with a supercell of 215 atoms containing a monovacancy at a) 1300 K, b) 1400 K, c) 1500K, d) 1600 K, e) 1700 K, and f) 1800 K. The phonons are obtained from the converged force constant using the s-TDEP method. The folded and unfolded phonon dispersions are shown by gray and a color map that indicates the spectral function of the unfolding, respectively.} 
\label{fig:Zr_215}
\end{figure}

\begin{figure}[H]\includegraphics[width=1.0\textwidth]{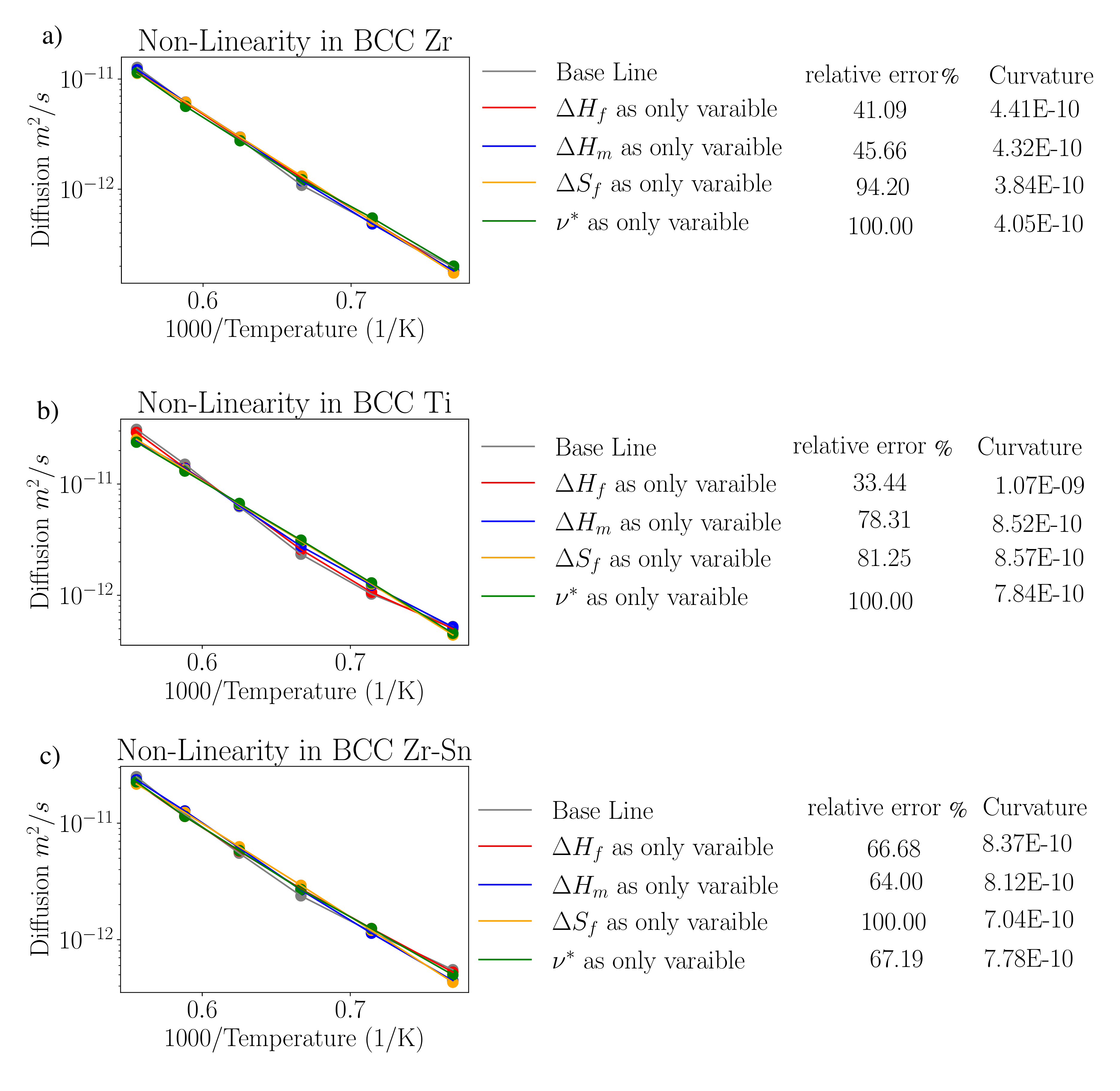}\hfill
\caption{The effect of temperature-dependence of each diffusion parameter ($\Delta H_f$, $\Delta H_m$, $\Delta S_f$, and $\nu^*$) on the curvature of the Arrhenius plots of diffusivity for a) bcc Zr, b) bcc Ti, and c) bcc Zr-0.46at\%Sn. Baseline is when the temperature-dependence of all the parameters are included. Other lines represent the Arrhenius plot when only one parameter is temperature-dependent and other parameters are fixed at their average values over temperature. The relative error is calculated as the summation of absolute differences with respect to the base line at each temperature. The curvature is calculated by fitting the diffusion curve to a second order polynomial.} 
\label{fig:non_linearity}
\end{figure}

\begin{figure}[H]\includegraphics[width=1.0\textwidth]{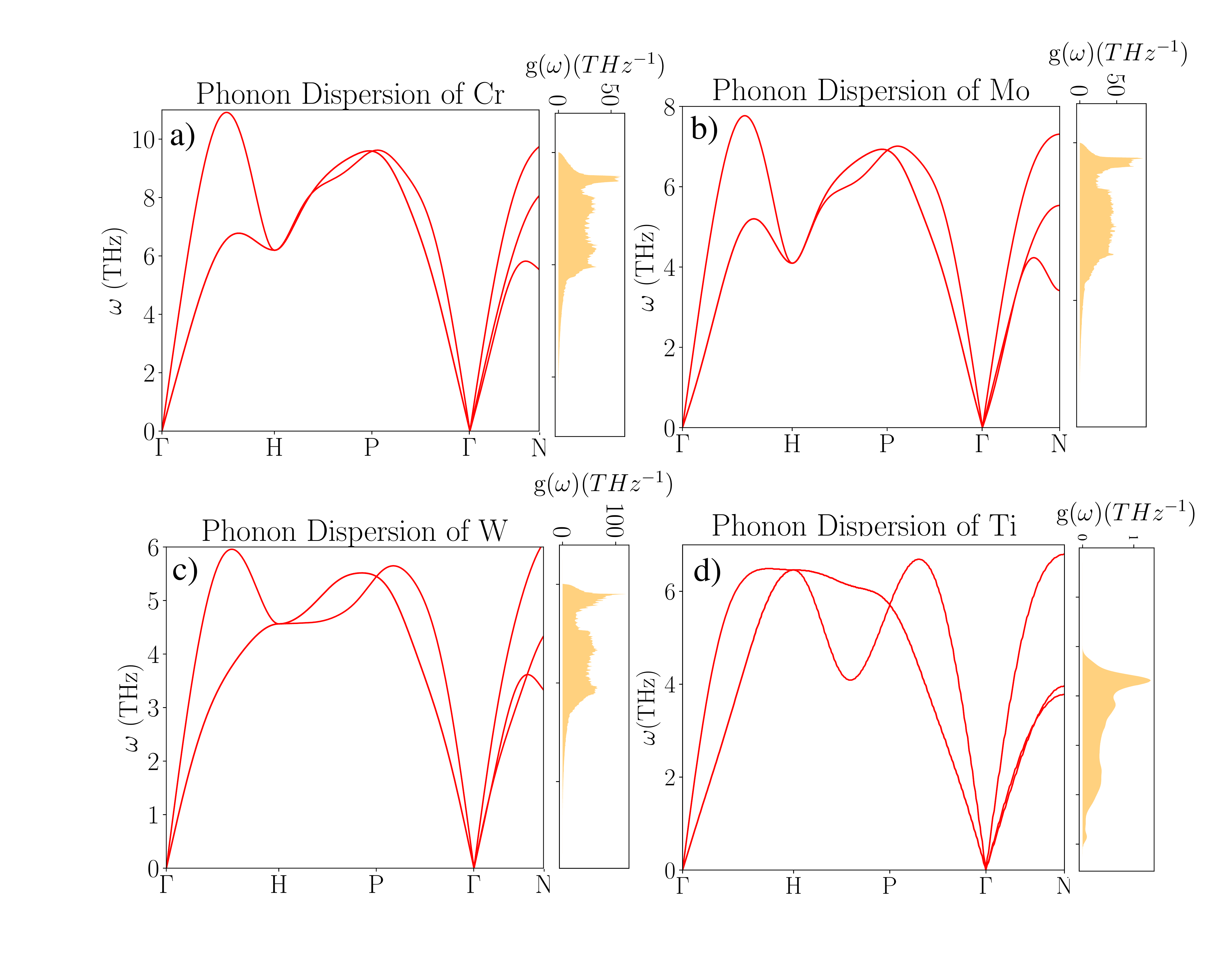}\hfill
\caption{Phonon dispersion and density of states for the defect-free bulk a) bcc Cr, b) bcc Mo, c) bcc W and d) bcc Ti (at 1400K). For bcc Ti, phonon softening along the $\left<111\right>$ phonon branch manifests itself as a dip in frequency located at $\frac{2}{3}L\left<111\right>$ (or $\frac{2}{3}$ of $\Gamma-P-H$ branch). The mechanically stable phases of bcc Cr, Mo, and W indicate no similar phonon softening effect.} 
\label{fig:allDispersion}
\end{figure}

\bibliographystyle{unsrt}

\end{document}